\documentclass[11pt]{article}   
\usepackage{latexsym}
\usepackage{bbm}      
\makeatletter
\def\@begintheorem#1#2{\trivlist%
 \item[\hskip \labelsep{\sffamily\bfseries #2\ #1}]\itshape}
\newtheorem{teo}{Theorem}[section]
\newtheorem{defi}[teo]{Definition}
\newtheorem{cor}[teo]{Corollary}
\newtheorem{lem}[teo]{Lemma}
\newtheorem{pro}[teo]{Proposition}
\newtheorem{_rem}[teo]{Remark}
\newtheorem{_eje}[teo]{Example}
\newenvironment{rem}{\def\@begintheorem##1##2{\trivlist%
 \item[\hskip\labelsep{\sffamily\bfseries ##2\ ##1}]}\begin{_rem}}{\end{_rem}}

\makeatother

\newenvironment{beweis}{{\em Proof:}}{\hfill $\rule{2mm}{2mm}$
\vspace{3mm}

}

\DeclareMathAlphabet{\Ma}{U}{msa}{m}{n}
\DeclareMathAlphabet{\Mb}{U}{msb}{m}{n}
\DeclareMathAlphabet{\Meuf}{U}{euf}{m}{n}

\def\got#1{\Meuf{#1}}
\def\wz{\widetilde{Z}}

\def\ev{\mbox{\tiny {\rm even}}}
\def\od{\mbox{\tiny {\rm odd}}}

\DeclareSymbolFont{ASMa}{U}{msa}{m}{n}
\DeclareSymbolFont{ASMb}{U}{msb}{m}{n}
\DeclareMathSymbol{\hrist}{\mathord}{ASMa}{"16}
\DeclareMathSymbol{\varkappa}{\mathalpha}{ASMb}{"7B}
\DeclareMathSymbol{\CrPr}{\mathord}{ASMb}{"6F}

\def\1{\mathbbm 1}
\def\EINS{\1}

  \def\al #1.{{\mathcal{#1}}}
  \def\ot #1.{{\got{#1}}}
  \def\C{\Mb{C}}
  \def\N{\Mb{N}}
  \def\R{\Mb{R}}
  \def\Sfin{S_\mathrm{\mbox{\tiny fin}}}
\title{\bf Twisted duality of the CAR-Algebra}
\author{
 {\sc Hellmut Baumg\"artel, Matthias Jurke}  \\[2mm] 
 {\footnotesize Mathematical Institute, University of Potsdam,}    \\   
 {\footnotesize Am Neuen Palais 10, Postfach 601~553,}             \\ 
 {\footnotesize D-14415 Potsdam, Germany.}                        \\[1mm]
 {\footnotesize baumg@rz.uni-potsdam.de}                           \\
\and
 {\sc Fernando Lled\'o}  \\[2mm] 
 {\footnotesize Institute for Pure and Applied Mathematics,}       \\
 {\footnotesize RWTH-Aachen, Templergraben 55,}                   \\ 
 {\footnotesize D-52062 Aachen, Germany.}                         \\[1mm]
 {\footnotesize lledo@iram.rwth-aachen.de}}
\date{\today{}}

\begin{document}
\maketitle
\begin{abstract} 
We give a complete proof of the twisted duality property 
$\al M.(\ot q.)'=\widetilde{Z}\,\al M.(\ot q.^\perp)\,\widetilde{Z}^*$
of the (self-dual) CAR-Algebra in any Fock representation.
The proof is based on the natural Halmos decomposition of the (reference)
Hilbert space when two suitable closed subspaces have been
distinguished. We use modular theory and techniques developed by 
Kato concerning pairs of projections in some essential steps of 
the proof. 

As a byproduct of the proof we obtain an explicit and simple formula for 
the graph of the modular operator. This formula can be also applied 
to fermionic free nets, hence giving a formula of the modular operator
for any double cone.
\end{abstract}

AMS-class: 46L10, 47A05, 81T05

%%%%%%%%%%%%%%%%%%%%%%%%%%%%%%%%%%%%%%%%%%%%%%%%%%%%%%%%%%%%%%%%%%%%%%%%%%%%%%
%%%%%%%%%%%%%%%%%%%%%%%%%%%%%%%%%%%%%%%%%%%%%%%%%%%%%%%%%%%%%%%%%%%%%%%%%%%%%%
\section{Introduction}

Twisted duality is a structural property of the von Neumann algebra
obtained from the CAR-Algebra (which is an abstract C*-algebra)
in any Fock representation. The
(self-dual) CAR-Algebra is uniquely given once a separable Hilbert 
space $\ot h.$ and an anti-unitary involution $\Gamma$ are specified
\cite{Araki70/71,ArakiIn87}. Now for any $\Gamma$-invariant subspace
$\ot q.$ of $\ot h.$ and any Fock state characterized by a so-called
basis projection $P$ we can canonically construct a von Neumann algebra
$\al M.(\ot q.)$. Twisted duality means that the equation
\begin{equation}\label{Td}
\al M.(\ot q.)'= \widetilde{Z}\,\al M.(\ot q.^\perp)\, \widetilde{Z}^*
\end{equation}
holds, where $\widetilde{Z}$ is 
a certain unitary twist operator to be introduced in the
following section and $\ot q.\oplus\ot q.^\perp=\ot h.$. Thus in order to
formulate duality in the context of the (self-dual) 
CAR-algebra one needs to distinguish two
closed subspaces $\ot q.$ and $\ot p.:=P\ot h.$ in the reference Hilbert
space $\ot h.$. The study of two closed subspaces of a Hilbert space
has a long and interesting history in functional analysis
(e.g.~\cite{Dixmier48,Davis58,Halmos69,Rieffel77,Avron94,Borac95,bKato95}) 
as well as applications in mathematical physics 
\cite{Avron94a}. 

We will see how the analysis of the relative position of these subspaces
will naturally suggest the strategy of the complete proof of 
eq.~(\ref{Td}) that we present in this paper.
Concretely, given $\ot p.$ and $\ot q.$ as before
we can canonically consider the Halmos decomposition 
$\ot h.=\ot h._0\oplus \ot h._1$ (cf.~\cite{Halmos69}), 
where $\ot h._0$ is the maximal subspace on which the 
orthoprojections corresponding to the closed subspaces
commute \cite[Section~III]{Borac95}. Our proof of (\ref{Td})
in the general case is based on the corresponding property for
the generic position situation where $\ot h._0=\{0\}$. The case
where $\ot p.$ and $\ot q.$ are in generic position 
allows to use modular theory \cite{bKadisonII,bBratteli87}
as well as results of Kato \cite{bKato95} for pairs of projections.
In this context we will characterize
the bicontinuity of different mappings that naturally appear here,
e.g.~the Tomita operator restricted to the one-particle 
Hilbert space. Further, we discuss systematically the relation of an 
important mapping $\varphi$ (and the components of its polar decomposition) 
used by Araki and Dell'Antonio \cite{Araki63,DellAntonio68}, to the modular
objects given in our case. The mapping $\varphi$ is introduced by these
authors to study the type of certain local von Neumann algebras.

The equation (\ref{Td}) appears naturally in the context of
algebraic quantum field theory, in particular 
in relation with Haag duality, which is one of its central concepts
(see e.g.~\cite{DHR71,DHR74,Summers82,bHaag92,bBaumgaertel95}).
Haag duality is a strengthening of
Einstein causality for a net of von Neumann algebras indexed by suitable  
regions in $\R^4$. In the context of the CCR-Algebra
(bosonic systems), one usually proceeds in two steps 
in order to prove this property
\cite{Araki63,Eckmann73,Osterwalder73,pLeyland78,Hislop86}: 
first one shows the so-called {\it abstract} duality
\[
{\cal M}({\got m}^{\perp_\sigma})={\cal M}({\got m})'\,,
\]
where ${\cal M}({\got m})'$ is the commutant of the von Neumann algebra 
${\cal M}({\got m})$ which is generated by the Weyl operators associated 
to a closed real subspace $\ot m.$ of the one-particle Hilbert space and 
${\got m}^{\perp_\sigma}$ denotes the symplectic complement of
${\got m}$. This result is then crucially used in a second step in 
order to reduce the proof of Haag duality to the 
discussion of certain real subspaces $\ot m.(\al O.)$ 
associated to sufficiently regular regions $\al O.$ in $\R^4$. 
In the context of the CAR-Algebra
(fermionic systems), and taking into account 
that now the generators of the algebra will anti-commute 
if the corresponding elements of the reference 
space are mutually orthogonal, one can adapt the notions of  
duality (cf.~(\ref{Td})) and Haag duality. 
(In the following we will avoid the use of the adjective `abstract').
Twisted duality (\ref{Td}) is mentioned (without proof) in 
\cite[Remark~4.9]{Araki70/71} and proved in \cite{Foit83}.
For the special case of the generic position situation see also
\cite[p.~496]{Wassermann98}. An important 
difference between the proof we present here and 
those in \cite{Foit83,Wassermann98} is that we will 
use the self-dual approach to the CAR-Algebra 
\cite{Araki70/71,ArakiIn87} and will consistently 
work with {\it complex} Hilbert subspaces (for further 
details about the relation between our proof and those in
\cite{Foit83,Wassermann98} we refer to 
Remark~\ref{rel1} and to Section~\ref{rel2}). 
This is not only a matter of elegance, but only the 
{\it explicit} use of a basis projection $P$ in order to specify 
the Fock states of the CAR-Algebra will allow to consider 
the natural Halmos decomposition of the reference Hilbert space. 
Therefore the whole strategy of the proof as well as various formulas
we prove in Section~\ref{Pairs} (e.g.~a simple and explicit expression
for the graph of the modular operator as well as a formula for the 
modular conjugation, cf.~Theorem~\ref{beta*beta} and Remark~\ref{rel1})
will depend on this choice.
Finally, the twisted duality
property (\ref{Td}) of the CAR-Algebra can be applied to the 
fermionic free nets defined in 
\cite{Lledo95,tJurke97,Lledo01} in order to prove Haag duality for 
these models. In the mentioned references the authors
present a direct way to construct nets of local C*- or 
von Neumann-algebras associated to massive (resp.~massless) models
for any half-integer spin (resp.~helicity) value. The nets for these 
models are naturally characterized by a 
net of local $\Gamma$-invariant linear subspaces 
$\al O.\mapsto\ot q.(\al O.)$ of the corresponding reference 
Hilbert space. Now the formulas for the graph of the modular operator
and modular conjugation (cf.~Theorem~\ref{beta*beta} and Remark~\ref{rel1})
can also be applied to the localized CAR-algebras associated with
$\ot q.(\al O.)$, where $\al O.$ is a double cone. 
We will give in Theorem~\ref{O} the corresponding 
formulas for the localized modular objects.

The paper is structured in 7 sections: 
In Section~\ref{CAR} we will state basic results concerning the CAR-Algebra 
that will be used later on. In the following section we will consider the
Halmos decomposition of a Hilbert space and state necessary and sufficient
conditions on the subspaces $\ot p.$ and $\ot q.$ in order that the
modular theory is well defined for $\al M.(\ot q.)$ and its Fock
vacuum vector $\Omega$.
In Section~\ref{Pairs} we will systematically analyze 
the context defined by two projections: 
the first one being the orthoprojection $Q$ onto $\ot q.$ 
and the other one being the basis projection $P$. 
The main goal here is to relate the objects that appear 
in the polar decomposition of the mapping $\varphi$ and of the Tomita
operator $S$ corresponding to $(\al M.(\ot q.),\Omega)$
and restricted to the one-particle space $\ot p.$. 
On the way to this goal we will give simple formulas
for the graphs of $\varphi^*\varphi$ and the modular operator
which show a beautiful symmetry w.r.t.~the interchange
$P\leftrightarrow Q$ (cf.~Theorem~\ref{beta*beta} and 
Proposition~\ref{Provarphi*varphi}).
These results will be applied in the next section where 
(\ref{Td}) is proved in the case where $\ot p.$ and $\ot q.$
are in generic position. In Section~\ref{rel2} the relation 
of the self-dual approach to the real subspace approaches 
in \cite{Foit83,Wassermann98} are pointed out. In the last section we
give a complete proof of (\ref{Td}) in the most general situation,
i.e.~for any $\Gamma$-invariant closed subspace $\ot q.$
and any Fock state. The proof is based
on the results of the previous two sections.

%%%%%%%%%%%%%%%%%%%%%%%%%%%%%%%%%%%%%%%%%%%%%%%%%%%%%%%%%%%%%%%%%%%%%%%%%%%%%
%%%%%%%%%%%%%%%%%%%%%%%%%%%%%%%%%%%%%%%%%%%%%%%%%%%%%%%%%%%%%%%%%%%%%%%%%%%%%
\section{Basic structure of the CAR-Algebra}\label{CAR}

In order to establish our notation we will begin this section 
collecting some standard results concerning 
the CAR-Algebra that will be needed later on.
For proofs and further results we refer to \cite{Araki70/71,ArakiIn87}.
In the following subsections we will also
consider additional structure of this
algebra necessary for our proof of twisted duality.

\begin{teo}\label{UniqueCAR}
Let $\ot h.$ be a complex Hilbert space with scalar product 
$\langle\cdot,\cdot\rangle$ and let $\Gamma$ be an anti-unitary
involution on it, i.e.~$\langle\Gamma f,\Gamma h\rangle=
\langle h,f \rangle$, for all $f,h\in \ot h.$. Then
{\rm CAR}$(\ot h.,\Gamma)$ denotes the algebraically unique 
{\rm C}*-algebra generated by $\EINS$ and $a(f)$, $f\in\ot h.$,
such that the following relations hold:
\begin{enumerate}
\renewcommand{\labelenumi}{(\roman{enumi})}
\item The mapping ${\got h} \ni f \mapsto a(f)$ is antilinear.
\item For any $f\in{\got h}$ one has $a(f)^* = a(\Gamma f)$.
\item For any $f, h \in {\got h}$ the equation $a(f) a(h)^* + a(h)^* a(f) 
= \langle f, h\rangle \,\EINS$ holds. 
\end{enumerate}
\end{teo}

Next we define a class of pure states of the preceding C*-algebra. An
orthoprojection $P$ of $\ot h.$ is called a {\it basis projection} if 
it satisfies the relation 
$P+\Gamma P \Gamma = \EINS$. These projections uniquely characterize
so-called {\it Fock states} $\omega_P$ by means of the equation
\begin{equation}\label{UniqueFock}
 \omega_P\Big( a(f)^*a(f) \Big):=0 \quad \mbox{for~any} 
                       \quad f\in\ot h.\quad\mbox{with}\quad  Pf=f \,.
\end{equation}
The antisymmetric Fock space is given by 
\begin{equation}\label{FockSpace}
 \ot F.:= \mathop{\oplus}\limits_{n=0}^\infty 
          \Big(\mathop{\land}\limits^n P\ot h.\Big)\,.
\end{equation}
In order to specify the Fock representation $\pi(a(f))$ of the
generators $a(f)$ we need to introduce the usual annihilation and creation
operators on $\ot F.$.

\begin{eqnarray*}
c(p)\,\Omega    &:=& 0\,,\\
c(p)\,(p_1 \land \ldots \land p_n) 
                &:=& \sum\limits_{r=1}^{n}\, (-1)^{r-1}\, 
              \langle p,p_r\rangle_{\got h} \;p_1 
              \land\ldots\land\widehat p_r \land \ldots \land p_n\,,\\[2mm]
c(p)^*\,\Omega  &:=& p\,,\\
c(p)^*\,(p_1 \land \ldots \land p_n) 
                &:=& p \land p_1 \land \ldots \land p_n\,,
\end{eqnarray*}
where $\Omega$ is the Fock vacuum in the subspace corresponding to
$n=0$ in the definition (\ref{FockSpace}) and $p,p_1,\ldots, p_n\in
P\ot h.$. $\widehat p_r$ means that the vector $p_r$ is omitted
in the wedge product. Finally, the Fock representation $\pi$
is defined by
\[
 \pi(a(f)):= c(P\Gamma f)^*+c(Pf)\,,\quad f\in \ot h.\,.
\]

In the rest of the paper we assume that a basis projection $P$ 
is given and when no confusion arises we will also simply write $a(f)$ 
instead of $\pi(a(f))$.
To prove twisted duality in Section~\ref{AbsDua}
we will need an explicit formula for the vector 
$a(f_n)\cdot\ldots\cdot a(f_1)\,\Omega$. Let $n,k,p$ be natural numbers
with $2p+k=n$ and define the following subset of the symmetric group
$\ot S._n$:
\[
\begin{array}{l} \ot S._{n,\,p}:=\left\{\left( {\begin{array}{cccccccc}
       n & n-1     &\cdots & n-2p+2  & n-2p+1 & k   & \cdots & 1 \\
\alpha_1 &\beta_1 &\cdots &\alpha_p &\beta_p & j_1 & \ldots & j_k   
               \end{array}}\right)\in \ot S._n\; \right. \\[5mm] 
 \;\;\;\;\;\;\;\;\;\;\;\;\;\;\;\;\; \left. \Big|\;\;
 \alpha_1>\ldots>\alpha_p\,,\; \alpha_l>\beta_l\,,\;l=1,\ldots, p 
 \quad\mbox{and}\quad n\geq j_1>j_2>\ldots j_k\geq 1
 \right\}\,.\end{array}
\]
Note that $\ot S._{n,\,p}$ contains $\left(\kern-1.5mm\begin{array}{c}
n \\n-2p \end{array}\kern-1.5mm\right) {\displaystyle \frac{(2p)!}{
 p!\,2^p}}\;$ elements.

\begin{pro}\label{Formel}
For $f_1,\ldots,f_n\in \ot h.$ the equation
\[
 \Big(a(f_n)\cdot\ldots\cdot a(f_1)\Big)\,\Omega = 
  \sum\limits_{\mbox{\tiny $\begin{array}{c}\pi\in\ot S._{n,\,p}
               \\[1mm] 0\leq 2p\leq n\end{array}$}}
  \!\!\!\!({\rm sgn}\,\pi)\;
  \prod\limits_{l=1}^p \;\langle Pf_{\alpha_l}\,,\,P\Gamma f_{\beta_l}\rangle
  \, P\Gamma f_{j_1} \land \ldots \land P\Gamma f_{j_k} 
\]
holds, where the indices $\alpha_l, \beta_l,j_1,\ldots,j_k$ are given
in the definition of $\ot S._{n,\,p}$ and where for $n=2p$ in the preceding
sum one replaces the wedge product by the vacuum $\Omega$.
\end{pro}
\begin{beweis}
See appendix.
\end{beweis}

Let $Z$ be the implementation on $\ot F.$ of the even-oddness automorphism
associated to the Bogoljubov unitarity $-\EINS$ \cite[p.~76]{ArakiIn87}.
It satisfies $Z=Z^*=Z^{-1}$ and therefore its spectral decomposition is
simply given by
\begin{equation}\label{SpectralZ}
Z=E^+ - E^- \,.
\end{equation}
Let further $X=X_{\ev}+X_{\od}$ be the
unique decomposition of any $X\in\mathrm{CAR}(\ot h.,\Gamma)$ into
its even and odd parts. The following result will be used in
Section~\ref{AbsDuaII}. 
\begin{lem}\label{EODecomp} 
Let $\pi$ be a Fock representation of $\mathrm{CAR}(\ot h.,\Gamma)$
and $Z=E^+ - E^-$ as before. Then for any $X\in\mathrm{CAR}(\ot h.,
\Gamma)$ we have
\begin{eqnarray*}
E^+\pi(X_{\ev})E^-&=&E^-\pi(X_{\ev})E^+=0\\
E^+\pi(X_{\od})E^+&=&E^-\pi(X_{\od})E^- =0\,.
\end{eqnarray*}
\end{lem}
\begin{beweis}
Recall that $Z_0\pi(X_{\ev})=\pi(X_{\ev})Z_0$. Multiplying from the left
by $E^+$ and from the right by $E^-$ we get
\[
 E^+\pi(X_{\ev})E^-=-E^+\pi(X_{\ev})E^-\,,
\]
which implies the first two equations. Similarly we obtain the  
equations corresponding to the odd part.
\end{beweis}
%%%%%%%%%%%%%%%%%%%%%%%%%%%%%%%%%%%%%%%%%%%%%%%%%%%%%%%%%%%%%%%%%%%%%%%%%%%%%%
\subsection{Direct sums and tensor products}

Let $(\ot h._k,\Gamma_k)$, $k=0,1$, be two Hilbert spaces with
anti-unitary involutions and let $P_k$ be two basis
projections for the corresponding spaces. We denote  
by $\pi_k$ the Fock representations of 
$\mathrm{CAR}(\ot h._k,\Gamma_k)$ on the 
antisymmetric Fock spaces $\ot F._k$. 
$Z_k$ are the implementers of the respective even-oddness automorphisms,
$k=0,1$. Motivated by \cite[p.~82]{DellAntonio68} 
(cf.~also \cite[p.~219]{bEvans98}) we consider:
\begin{pro}\label{Tensor}
With the preceding notation put 
$\ot h.:=\ot h._0\oplus\ot h._1$ and $\Gamma:=
\Gamma_0\oplus\Gamma_1$ and consider $\mathrm{CAR}(\ot h.,\Gamma)$.
The representations defined on $\ot F._0\otimes\ot F._1$ by 
\begin{eqnarray}
 \label{DefTensor}
 \pi\Big(a(f_0\oplus f_1)\Big)&:=&
   \pi_0\Big(a(f_0)\Big)\otimes \EINS +
    Z_0\otimes\pi_1\Big(a(f_1)\Big)             \\
   \label{DefTensor2}
 \pi\Big(a(f_0\oplus f_1)\Big)&:=&
   \pi_0\Big(a(f_0)\Big)\otimes Z_1 +
    \EINS\otimes\pi_1\Big(a(f_1)\Big)\,,
   \;\;f_k\in\ot h._k\,,k=0,1\,,
\end{eqnarray}
are the Fock representations of $\mathrm{CAR}(\ot h.,\Gamma)$ w.r.t.~the
basis projection $P:=P_0\oplus P_1$ on $(\ot h.,\Gamma)$. Further,
the even-oddness automorphism of $\mathrm{CAR}(\ot h.,\Gamma)$
is implemented on $\ot F._0\otimes\ot F._1$ by $Z:=Z_0\otimes Z_1$.
\end{pro}
\begin{beweis}
We check first that the definition (\ref{DefTensor}) is compatible
with the algebraic structure of $\mathrm{CAR}(\ot h.,\Gamma)$.
Indeed, for any $f_k,h_k \in\ot h._k$, $k=0,1$,
\[
 \pi\Big(a(f_0\oplus f_1)\Big)^*=
   \pi_0\Big(a(\Gamma_0 f_0)\Big)\otimes \EINS +
    Z_0\otimes\pi_1\Big(a(\Gamma_1f_1)\Big)
  = \pi\Big(a(\Gamma(f_0\oplus f_1))\Big)
\]
and 
\begin{eqnarray*}
\lefteqn{\!\!\!\!\!\!\!\!\!\!\!\!\!\!\!\!\!\!\!\!\!\!\!\!\!\!\!\!\!\!\!\!\!\!
 \pi\Big(a(f_0\oplus f_1)\Big)\pi\Big(a(h_0\oplus h_1)\Big)^*
 +\pi\Big(a(h_0\oplus h_1)\Big)^*\pi\Big(a(f_0\oplus f_1)\Big)} \\[2mm]
   &=& \pi_0\Big(a(f_0)a(h_0)^*\Big)\otimes \EINS +
       \EINS\otimes\pi_1\Big(a(f_1)a(h_1)^*\Big)\\[2mm]
   & & + \pi_0\Big(a(h_0)^*a(f_0)\Big)\otimes \EINS +
       \EINS\otimes\pi_1\Big(a(h_1)^*a(f_1)\Big)\\[2mm]
   & & + \underbrace{
         Z_0\pi_0\Big(a(h_0)^*\Big)\otimes\pi_1\Big(a(f_1)\Big)
       + \pi_0\Big(a(h_0)^*\Big)Z_0\otimes\pi_1\Big(a(f_1)\Big)}_{=0}\\
   & & + \underbrace{
         Z_0\pi_0\Big(a(f_0)\Big)\otimes\pi_1\Big(a(h_1)^*\Big)
       + \pi_0\Big(a(f_0)\Big)Z_0\otimes\pi_1\Big(a(h_1)^*\Big)}_{=0}\\[2mm]
   &=& \langle f_0\oplus f_1 , h_0\oplus h_1\rangle \,\EINS\otimes\EINS\,.
\end{eqnarray*}
Consider next the vector state associated to $\Omega_0\otimes\Omega_1$,
where $\Omega_k$ are the Fock vacua in $\ot F._k$, $k=0,1$. In this case
we have
\begin{eqnarray*}
\lefteqn{\!\!\!\!\!\!\!\!\!\!\!\!\!\!\!\!\!\!\!\!\!\!\!\!\!\!\!\!\!\!\!\!\!\!
\Big\langle \Omega_0\otimes\Omega_1,
   \pi\Big(a(f_0\oplus f_1)\Big)^*\pi\Big(a(f_0\oplus f_1)\Big)
   \Omega_0\otimes\Omega_1\Big\rangle} \\ 
   &=& \Big\|\pi_0\Big(a(f_0)\Big)\Omega_0\otimes\Omega_1
      +\Omega_0\otimes \pi_1\Big(a(f_1)\Big)\Omega_1 \Big\|^2 \\[2mm]
    &=& \|P_0\Gamma_0 f_0\|^2+\|P_1\Gamma_1 f_1\|^2
       = \|(\EINS-P) (f_0\oplus f_1)\|^2\,.
\end{eqnarray*}
Thus by the uniqueness of the characterizing condition (\ref{UniqueFock})
of a Fock state, together with the fact that the vectors 
$\pi\Big(a(f_0\oplus f_1)\Big)\Omega_0\otimes\Omega_1$, $f_k\in\ot h._k$,
generate the whole space $\ot F._0\otimes\ot F._1$,
we have that $\pi$ is actually the Fock representation
of $\mathrm{CAR}(\ot h.,\Gamma)$ w.r.t.~the basis projection
$P$ with Fock vacuum $\Omega_0\otimes\Omega_1$.

We still need to show that $Z_0\otimes Z_1$ implements the even-oddness
automorphism of $\mathrm{CAR}(\ot h.,\Gamma)$ on $\ot F._0\otimes
\ot F._1$. Note that $\alpha:=\mathrm{ad}(Z_0\otimes Z_1)$
satisfies $\alpha^2=\mathrm{id}$ and it is enough to consider the 
action of $\alpha$ on the generators:
\[
(Z_0\otimes Z_1)\pi\Big(a(f_0\oplus f_1)\Big)(Z_0\otimes Z_1)
 = Z_0\pi_0\Big(a(f_0)\Big)Z_0 \otimes Z_1^2 +
    Z_0^3\otimes Z_1\pi_1\Big(a(f_1)\Big)Z_1
 = \pi\Big(-a(f_0\oplus f_1)\Big)\,.
\]
Eq.~(\ref{DefTensor2}) is shown similarly.
\end{beweis}
\begin{cor}\label{EOTensor}
For $X_0\in\mathrm{CAR}(\ot h._0,\Gamma_0)$ and $X_1=X_{1,\ev}+X_{1,\od}
\in\mathrm{CAR}(\ot h._1,\Gamma_1)$ we have
\begin{eqnarray*} 
 \pi(X_0)&=&\pi_0(X_0)\otimes\EINS\quad\mathrm{and}\quad \\
 \pi(X_1)&=&\EINS\otimes\pi_1(X_{1,\ev})+Z_0\otimes\pi_1(X_{1,\od})\\
          &=&E_0^+\otimes\pi_1(X_1)+E_0^-\otimes Z_1\pi_1(X_{1,\od})Z_1\,,
\end{eqnarray*}
where $Z_0=E_0^+-E_0^-$.
\end{cor}

%%%%%%%%%%%%%%%%%%%%%%%%%%%%%%%%%%%%%%%%%%%%%%%%%%%%%%%%%%%%%%%%%%%%%%%%%%%%%
\subsection{The subspace $\ot q.$ and twisted causality}\label{Q}

Let $(\ot h.,\Gamma)$ be as in Theorem~\ref{UniqueCAR} 
and denote by $\ot q.$ a closed
$\Gamma$-invariant subspace of $\ot h.$. We can naturally associate
with the subspace $\ot q.$ a von Neumann algebra that acts on the 
antisymmetric Fock space characterized by the basis projection $P$:
\begin{equation}\label{VNAm}
 \al M.(\ot q.):= \Big( \{ a(q)\mid q\in \ot q.\}\Big)''
                  \subset \al L.(\ot F.)\,.
\end{equation}

In order to be able to formulate causality or 
duality in the context of the CAR-Algebra it is necessary to 
introduce a so-called {\em twist operator} $\widetilde{Z}$
on $\ot F.$. Indeed, this operator allows to express orthogonality
relations of subspaces of $\ot h.$ in terms of the commutant of a 
suitable von Neumann algebra. 
Recalling the definition of $Z$ before Eq.~(\ref{SpectralZ}) we
define the twist operator as usual \cite[Eq.~(26)]{Bisognano76}
\begin{equation}\label{twist}
 \widetilde{Z} := \frac{1}{1+i}(\EINS+iZ)\quad\mathrm{and}\quad
 \wz a(f)\wz^* =iZ a(f)\,,\;f\in\ot h.\,.
\end{equation}
Putting
\[
 \eta(n):= \left\{ \begin{array}{rlcl}
           1, &\mbox{if}& n & \mbox{even}\,,\\
           -i,&\mbox{if}& n & \mbox{odd}\,,
\end{array} \right.
\]
we may describe the action of $\wz$ and ${\wz}^*$ on $\ot F.$ by 
$\widetilde{Z}(p_1 \land \ldots \land p_n)= \eta(n)\,
p_1 \land \ldots \land p_n$ and $\widetilde{Z}^*(p_1 \land \ldots \land p_n)= 
\overline{\eta(n)}\,p_1 \land \ldots \land p_n$.
Further, considering the spectral projections $E^\pm$ of $Z$ 
we also have the relation
\begin{equation}\label{DecompTwist}
\wz = E^+-iE^-\,.
\end{equation}
The algebra $\widetilde{Z}\,\al M.(\ot q.)\, \widetilde{Z}^*$ is 
usually called `twisted' algebra. It is now immediate to check the
following inclusion, which expresses the twisted causality property
in the present context.
\begin{pro}\label{TwisCau}
Let $\ot q.$ be a closed $\Gamma$-invariant subspace of $\ot h.$
as before. Then the inclusion 
$\widetilde{Z}\,\al M.(\ot q.^\perp)\, \widetilde{Z}^* 
\subseteq \al M.(\ot q.)'$ holds. 
\end{pro}
\begin{rem}
Twisted duality strengthens this relation by turning the preceding
inclusion into an equality for {\em any} closed $\Gamma$-invariant
subspace $\ot q.$ of $\ot h.$.
\end{rem}
\begin{rem}\label{EOGen}
Writing now explicitly the Fock representation and considering 
Eq.~(\ref{twist}) as well as the 
even-odd grading of the CAR-Algebra we have the following ways of
generating the twisted von Neumann algebra:
\begin{eqnarray*}
\widetilde{Z}\,\al M.(\ot q.^\perp)\, \widetilde{Z}^*
  &=& \Big\{ Z\pi(a(q^\perp))  \mid q^\perp \in\ot q.^\perp\Big\}''\\
  &=& \Big\{\pi(Y_{\ev})+ iZ\pi(Y_{\od}))  \mid 
      Y=Y_{\ev}+ Y_{\od}\in
        \mathrm{CAR}(\ot q.^\perp,\Gamma\hrist\ot q.^\perp) \Big\}''\,.
\end{eqnarray*}
\end{rem}
%%%%%%%%%%%%%%%%%%%%%%%%%%%%%%%%%%%%%%%%%%%%%%%%%%%%%%%%%%%%%%%%%%%%%%%%%%%%%%
%%%%%%%%%%%%%%%%%%%%%%%%%%%%%%%%%%%%%%%%%%%%%%%%%%%%%%%%%%%%%%%%%%%%%%%%%%%%%%
\section{The Halmos decomposition of a Hilbert space}\label{Sec3}

From the preceding subsection we see that to formulate the duality
property in the present context one needs to distinguish two closed
subspaces of the reference space $\ot h.$: the subspace $\ot p.$
(one-particle Hilbert space) which is given by the basis projection
$P$ and the $\Gamma$-invariant subspace $\ot q.$ to
which we associate the orthoprojection $Q$. Therefore it is natural
to consider the Halmos decomposition \cite{Halmos69} of $\ot h.$
w.r.t.~$\ot p.$ and $\ot q.$, which is given by
\begin{equation}\label{Halmos}
 \ot h.=\ot h._0\oplus \ot h._1\,,
\end{equation}
where $\ot h._0=(\ot p.\cap\ot q.)\oplus(\ot p.\cap\ot q.^\perp)\oplus
(\ot p.^\perp\cap\ot q.)\oplus(\ot p.^\perp\cap\ot q.^\perp)$ and
$\ot h._1=\ot h.\ominus \ot h._0$. $\ot h._0$ is actually the maximal
subspace where $P$ and $Q$ commute \cite[Section~III]{Borac95}.
In $\ot h._1$ the subspaces $\ot p.$ and $\ot q.$ are said to be in 
{\em generic position} \cite{Halmos69}, in the sense that the preceding
four mutual intersections of the subspaces 
$\ot p.$, $\ot q.$, $\ot p.^\perp$ and $\ot q.^\perp$ are 
equal to $\{0\}$.

In the present context the decomposition (\ref{Halmos}) is also natural
because it allows to separate the general situation into simpler `pieces'
on $\ot h._0$ and $\ot h._1$:

\begin{lem}\label{LemZerlegung}
Consider $(\ot h.,\Gamma)$ and $P$,$Q$ as before. Let further $R_0$
be the orthoprojection onto the subspace $\ot h._0$ in (\ref{Halmos}).
Then we have
\[
 R_0P=PR_0\;,\quad R_0Q=QR_0\quad\mathrm{and}\quad 
 R_0\Gamma=\Gamma R_0\,.
\]
\end{lem}
\begin{beweis}
The first two equations are clear from the form of $\ot h._0$ given
after (\ref{Halmos}) and recall that $R_0$ is the maximal 
orthoprojection with the property $R_0\, PQ=R_0\, QP$. Now from 
\[
 \Gamma R_0\Gamma\, PQ=\Gamma R_0\,(\EINS- P)Q\Gamma=
 \Gamma R_0\Gamma\, QP
\]
and the maximality of $R_0$ we must have $\Gamma R_0\Gamma\leq R_0$.
Finally, from the $\Gamma$-invariance of $\ot h.$ and since $\Gamma^2=
\EINS$ we get $R_0\Gamma=\Gamma R_0$.
\end{beweis}

\begin{rem}\label{RemZerlegung}
The preceding result allows to consider the following restrictions:
\[
 P_k:=P\hrist\ot h._k\;,\quad Q_k:=Q\hrist\ot h._k\;,\quad\mathrm{and}
 \quad \Gamma_k:=\Gamma\hrist\ot h._k\,,\;\;k=0,1\,.
\]
Further, $P_k$ is a basis projection on $(\ot h._k,\Gamma_k)$
characterizing the Fock representation $\pi_k$ of CAR$(\ot h._k,\Gamma_k)$,
$k=0,1$. This possibility to split off the general situation in an 
abelian piece and a generic position piece will be essential for the 
proof of our main theorem in Section~\ref{AbsDuaII}. 
In this context we will need to split again the abelian part 
as $\ot h._0=\ot h._{01}\oplus\ot h._{02}$, where
$\ot h._{01}:=(\ot p.\cap\ot q.^\perp)\oplus\Gamma(\ot p.\cap\ot q.^\perp)$
and $\ot h._{02}:=(\ot p.\cap\ot q.)\oplus\Gamma(\ot p.\cap\ot q.)$.
\end{rem}

%%%%%%%%%%%%%%%%%%%%%%%%%%%%%%%%%%%%%%%%%%%%%%%%%%%%%%%%%%%%%%%%%%%%%%%%%%%%%%
\subsection{Modular theory}\label{ModCAR}

Let $(\ot h.,\Gamma)$, $P$ and $\ot q.$ be given as in Subsection~\ref{Q} 
and denote by $\ot p.:=P\ot h.$ the corresponding one-particle Hilbert
space. We will give in this subsection necessary and sufficient conditions
on the subspaces $\ot p.$ and $\ot q.$ in order that the Fock vacuum 
$\Omega$ becomes a cyclic and separating vector for the von Neumann 
algebra $\al M.(\ot q.)$ (cf.~\cite[p.~234]{bEvans98}). 
We will see that in the case where the subspaces
$\ot p.$ and $\ot q.$ are in generic position we can use the results 
in modular theory \cite{bKadisonII,bBratteli87} for the pair 
$(\al M.(\ot q.),\Omega)$. These techniques will be essentially used in
the next two sections for the proof of the twisted duality 
property in the generic position context.

\begin{lem}\label{Lem.3.1}
Let $\ot q.\subset \ot h.$ be a closed $\Gamma$-invariant subspace. 
Then the following conditions are equivalent:
\begin{itemize}
\item[(i)] $q\in \ot q.$ and $Pq=0$ implies $q=0$.
\item[(ii)] $P(\ot q.^\perp)$ is a dense submanifold of $\ot p.$.
\item[(iii)] $\ot q.\cap \ot p.=\{ 0\}$.
\end{itemize}
\end{lem}
\begin{beweis}
To show (iii)$\Rightarrow$(i) take $q\in\ot q.$ with
$Pq=0$. From $\Gamma P\Gamma =\EINS-P$ and the 
$\Gamma$-invariance of $\ot q.$ we have that $\Gamma q\in
\ot q.\cap\ot p.=\{ 0\}$, thus $q=0$.
The other implications (i)$\Rightarrow$(ii)$\Rightarrow$(iii) are 
checked similarly.
\end{beweis}

\begin{pro}\label{Teo.3.2}
Let $\ot q.$ be a closed $\Gamma$-invariant subspace of $\ot h.$ as
before. Then we have:
\begin{itemize}
\item[(i)] The vacuum vector $\Omega$ is cyclic for $\al M.(\ot q.)\;$
  iff $\;P\ot q.$ is a dense submanifold of $\ot p.$. 
 \item[(ii)] The vacuum vector $\Omega$ is separating for $\al M.(\ot q.)\;$
  iff $\;P(\ot q.^\perp)$ is a dense submanifold of $\ot p.$. 
\end{itemize}
\end{pro}
\begin{beweis}
(i) We assume that $\Omega$ is cyclic for $\al M.(\ot q.)$ and let
$p\in \ot p.$ be a vector satisfying $p\perp P\ot q.$. From this,
from Proposition~\ref{Formel} and from the structure of the Fock
space $\ot F.$ (recall Eq.~(\ref{FockSpace})) we have
\begin{eqnarray*}
  p &\perp& {\rm span}\,\{a(q_1)\cdot\ldots\cdot a(q_n)\Omega\mid
             q_1,\ldots q_n \in\ot q.\,,\; n\in\N \}\;,\quad\mathrm{thus}\\
   p &\perp& \{A\Omega\mid A\in\al M.(\ot q.)\} \,.
\end{eqnarray*}
Now since $\Omega$ is cyclic for $\al M.(\ot q.)$ we must have 
$p=0$. 

Assume now that $P\ot q.\subset\ot p.$ is a dense submanifold, so that
$\mathop{\oplus}\limits_{n=0}^\infty 
 \Big(\mathop{\land}\limits^n P\ot q.\Big)\subset \ot F.$ is also
dense (here $\mathop{\oplus}\limits_{n=0}^\infty$ denotes the algebraic
direct sum). From Proposition~\ref{Formel} we obtain the inclusions
\[
 \mathop{\oplus}\limits_{n=0}^\infty 
 \Big(\mathop{\land}\limits^n P\ot q.\Big)
 \subset \al M.(\ot q.)\,\Omega \subset \ot F.\,,
\]
which imply that $\Omega$ is cyclic for $\al M.(\ot q.)$.

(ii) Suppose now that $\Omega$ is a separating vector for
$\al M.(\ot q.)$. We show that this implies part~(i)
of Lemma~\ref{Lem.3.1}. So, let $q\in\ot q.$ satisfy $Pq=0$. 
Since $0=Pq=a(\Gamma q)\Omega$ and since $\Omega$ is separating
for $\al M.(\ot q.)$ we must have $a(\Gamma q)=0$, which implies
$q=0$.

Finally, assume that $P(\ot q.^\perp)$ is a dense submanifold of 
$\ot p.$. By part (i) of the present theorem
applied to $\ot q.^\perp$, which is also $\Gamma$-invariant,
we know that $\Omega$ is cyclic for $\al M.(\ot q.^\perp)$.
Further, since $\widetilde{Z}\Omega=\Omega$ we have that $\Omega$ is also 
cyclic for $\widetilde{Z}\,\al M.(\ot q.^\perp)\, \widetilde{Z}^* $ and consequently 
by Proposition~\ref{TwisCau} also
for $\al M.(\ot q.)'\supseteq \widetilde{Z}\,\al M.(\ot q.^\perp)\, \widetilde{Z}^*$.
This shows that $\Omega$ is separating for $\al M.(\ot q.)$.
\end{beweis}

By the preceding result we know that if $\ot q.$ is a closed 
$\Gamma$-invariant subspace of $\ot h.$ where 
$P\ot q.$ as well as $P(\ot q.^\perp)$ are dense submanifolds
of $\ot p.$, then the modular theory is well-defined for the 
pairs $(\al M.(\ot q.),\Omega)$ and $(\al M.(\ot q.^\perp),\Omega)$.
Denote by $S$ and $T$ the Tomita operators
corresponding to these pairs respectively. We will next study 
their action on the submanifolds $P\ot q.$ and $P(\ot q.^\perp)$.

\begin{lem}\label{Lem.3.3}
With the preceding notation we have for  $q\in\ot q.$ 
and $q^\perp\in\ot q.^\perp$: 
\begin{itemize}
\item[(i)] $S(Pq)=P\Gamma q$.
\item[(ii)] $T(Pq^\perp) =
            P\Gamma q^\perp =-S^* (Pq^\perp)$.
\end{itemize}
\end{lem}
\begin{beweis}
The first two equations follow by direct computation:
\[
 S(Pq)=S\Big(a(\Gamma q)\,\Omega\Big)
  = a(\Gamma q)^*\,\Omega= a(q)\,\Omega=P\Gamma q\,,
\]
and similarly for $T$. To prove the last equation
recall that $S^*$ is actually the Tomita operator of 
$\al M.(\ot q.)'\supseteq \widetilde{Z}\,\al M.(\ot q.^\perp)\, \widetilde{Z}^*$. Note further
that
\[
 i\,\Big(\widetilde{Z} a(\Gamma q^\perp) \widetilde{Z}^*\Big)\,\Omega =i\, \widetilde{Z} a(\Gamma q^\perp)\,\Omega
  =i\,\left(\frac{1-i}{1+i}\right) Pq^\perp = Pq^\perp\,.
\]
From this we can finally check,
\[
 S^* (Pq^\perp)=-i\, S^*\Big( \widetilde{Z} a(\Gamma q^\perp) 
 \widetilde{Z}^*\,\Omega \Big)
 =-i\, \widetilde{Z}a(q^\perp)\widetilde{Z}^*\,\Omega=-P\Gamma q^\perp 
 =-T (Pq^\perp)\,,
\]
and the proof is concluded.
\end{beweis}

%%%%%%%%%%%%%%%%%%%%%%%%%%%%%%%%%%%%%%%%%%%%%%%%%%%%%%%%%%%%%%%%%%%%%%%%%%%%%%
%%%%%%%%%%%%%%%%%%%%%%%%%%%%%%%%%%%%%%%%%%%%%%%%%%%%%%%%%%%%%%%%%%%%%%%%%%%%%%
\section{Pairs of projections in generic position and the mapping $\varphi$}
\label{Pairs}

In this section we will consider the mathematically richest situation
which appears when the closed subspaces $\ot p.$ and $\ot q.$ 
are in generic position.
Let $P$ and $Q$ be the corresponding orthoprojections satisfying as usual
the relations $\Gamma P\Gamma =\EINS-P=P^\perp$ and $Q\Gamma=\Gamma Q$. 
Motivated by Proposition~\ref{Teo.3.2} we will also assume here that
\begin{equation}\label{Durch1}
\ot p.\cap\ot q.=\{0\}=\ot p.\cap\ot q.^\perp
           \;,\quad\mbox{where}\;\ot p.:=P\ot h.\,,
\end{equation}
which directly implies using the basis projection property:
\begin{equation}\label{Durch2}
\ot p^\perp.\cap\ot q.=\{0\}=\ot p.^\perp\cap\ot q.^\perp
           \;,\quad\mbox{where}\;\ot p.^\perp=P^\perp\ot h.\,.
\end{equation}
Thus in the notation of Section~\ref{Sec3} we have the extremal case 
where $\ot p.$ and $\ot q.$ are already in generic position and the
Halmos decomposition (\ref{Halmos}) is trivial in the sense
that $\ot h._0=\{0\}$.

Note that the properties like $\ot p.\,\cap\,\ot q.=\{0\}$ can be 
also expressed by the corresponding projections $P$ and $Q$, because
the orthoprojection onto the intersection 
$\ot p. \cap\ot q.$ is given by 
\[
    \mathop{{\rm s-}\lim}\limits_{n\to \infty}\,(PQ)^n
   =\mathop{{\rm s-}\lim}\limits_{n\to \infty}\,(QP)^n\,.
\]
Note further that by Lemma~\ref{Lem.3.1} the intersection assumptions
in (\ref{Durch1}) are equivalent to the density conditions in 
Proposition~\ref{Teo.3.2}.

\begin{rem}
The following useful density statements are immediate consequences 
of the assumption that $\ot p.$ and $\ot q.$ are in generic position.
If $\ot r.\subseteq \ot q.$ (or $\ot r.\subseteq \ot q.^\perp$)
is a dense linear submanifold in $\ot q.$ (resp.~in $\ot q.^\perp$),
then $P\ot r.$ is dense in $\ot p.$ and 
$P^\perp\ot r.$ is dense in $\ot p.^\perp$.
The same holds if $Q$ and $P$ are interchanged. Thus we have for
example that $Q\ot p.^\perp$ is dense in $\ot q.$, $PQ\ot p.$
is dense in $\ot p.$ etc.
\end{rem}

We will begin next a systematic analysis of a mapping $\varphi$
that can be naturally defined in the present context. Put
\begin{eqnarray*}
 \ot H._\varphi &:=&\Big\{ (q,q^\perp)\in\ot q.\times\ot q.^\perp \mid
                      P^\perp(q+q^\perp)=0\Big\} \\
\ot H._\rho     &:=&\Big\{ (q,q^\perp)\in\ot q.\times\ot q.^\perp \mid
                      Pq =Pq^\perp\Big\} \,.
\end{eqnarray*}
\begin{lem}\label{Rho-1}
The sets $\ot H._\varphi$ and $\ot H._\rho$ are graphs of linear,
injective and closed mappings $\varphi,\rho\colon \ot q.\to\ot q.^\perp$
with dense domains and dense images. The graphs $\,{\rm gra}\,\varphi=
\ot H._\varphi$ and $\,{\rm gra}\,\rho =\ot H._\rho $ can be parametrized
by $\ot p.$ resp.~$\ot p.^\perp$ as
\begin{eqnarray}
\label{GraPhi}
 {\rm gra}\,\varphi &:=& \Big\{(Qp,Q^\perp p) \mid p\in\ot p.\Big\} \\
 {\rm gra}\,\rho    &:=& \Big\{(Qp^\perp,-Q^\perp p^\perp) 
                         \mid p^\perp\in\ot p.^\perp\Big\}\,,\nonumber
\end{eqnarray}
where the domains and images are given explicitly. Moreover, the 
equation $\rho^{-1}=\varphi^*$ holds.
\end{lem}
\begin{beweis}
We consider first the mapping $\varphi$ and one can similarly argue
for $\rho$. First note that the assignment $\varphi(q):=q^\perp$
if $P^\perp (q+q^\perp)=0$ is a well-defined linear map. Indeed, if
$(q,q_1^\perp)\in\ot H._\varphi\ni(q,q_2^\perp)$, then
$q_1^\perp=q_2^\perp$, because in this case $q_1^\perp-q_2^\perp\in
\ot q.^\perp\cap\ot p.=\{0\}$. Injectivity is proved analogously.

Next we show Eq.~(\ref{GraPhi}). Let $\ot p.\ni p=Qp+Q^\perp p$, so 
that $P^\perp (Qp+Q^\perp p)=0$ and we have $Q\ot p.\subseteq 
\mathrm{dom}\,\varphi$ as well as $Q^\perp\ot p.\subseteq 
\mathrm{ima}\,\varphi$. To show the reverse inclusions take
$q_0\in\mathrm{dom}\,\varphi$, i.e.~$P^\perp (q_0+q_0^\perp)=0$
for some $q_0^\perp\in \ot q.^\perp$. But this implies that
$q_0= Q(q_0+q_0^\perp)\in Q\ot p.$ and $q_0^\perp= 
Q^\perp(q_0+q_0^\perp)\in Q^\perp\ot p.$ and Eq.~(\ref{GraPhi})
is proved. Note that by the preceding remark the domain and 
image of $\varphi$ are dense in $\ot q.$ resp.~$\ot q.^\perp$
and it is easy to see that $\mathrm{gra}\,\varphi$ is closed.

Finally, it remains to show that $\rho^{-1}=\varphi^*$. Recall 
first that
\[
 \mathrm{gra}\,\rho^{-1}=\Big\{(Q^\perp p^\perp,-Q p^\perp) 
                         \mid p^\perp\in\ot p.^\perp\Big\}\,.
\]
Using the isometric mapping $U\colon\ \ot q.\times \ot q.^\perp
\to \ot q.^\perp\times\ot q.$ given by $U(q,q^\perp):=(q^\perp,-q)$
we may use the well known formula (see \cite[p.~124]{bAchieser81})
\[
 \mathrm{gra}\,\varphi^*=(\ot q.^\perp\!\times\ot q.)\ominus
                         U(\mathrm{gra}\,\varphi)\,.
\]
Therefore $(q^\perp,q)\in \mathrm{gra}\,\varphi^*$ iff
$\Big\langle (q^\perp,q), (Q^\perp p,-Q p)\Big\rangle =0$ for all
$p\in\ot p.$ iff $\langle q^\perp, p\rangle=\langle q, p\rangle$
for all $p\in\ot p.$ iff $p^\perp:= q^\perp - q \in\ot p.^\perp$
iff $(q^\perp,q)\in\mathrm{gra}\,\rho^{-1}$.
\end{beweis}

\begin{rem}
 Note that the preceding lemma depends only on the assumption
 that $P$ and $Q$ are in generic position (the different role
 of $P$ and $Q$ w.r.t.~$\Gamma$, i.e.~$Q\Gamma =\Gamma Q$ and
 $\Gamma P\Gamma = P^\perp$, is not used). This means that
the preceding lemma remains true if we do the following replacements
\[
 Q\to P \qquad\mathrm{and} \qquad P\to Q^\perp \,.
\]
Then we obtain:
\end{rem}
\begin{cor}\label{Lambda}
 Put 
\[
 \ot H._\lambda := \Big\{ (p,p^\perp)\in\ot p.\times\ot p.^\perp \mid
                      Q^\perp p =Q^\perp p^\perp\Big\} \,.
\]
Then $\ot H._\lambda$ is the graph of a linear injective closed 
mapping $\lambda\colon\ \ot p.\to \ot p.^\perp$ with dense domain
and image. $\mathrm{gra}\,\lambda$ can be parametrized by $\ot q.$:
\[
\mathrm{gra}\,\lambda=\Big\{(P q,-P^\perp q) 
                      \mid q\in\ot q.\Big\}\,.
\]
\end{cor}

The parametrization of $\varphi$ and $\rho$ in terms of $\ot p.$ 
resp.~$\ot p.^\perp$ suggests to consider the following mappings:
\begin{eqnarray}
\label{1}&Q\phantom{^\perp}\colon& 
    \ot p.^\perp \longrightarrow  \ot q. \\
\label{2}&Q^\perp\colon& \ot p.^\perp \longrightarrow  \ot q.^\perp \\
\label{3}&Q\phantom{^\perp}\colon& 
     \ot p.\phantom{^\perp} \longrightarrow  \ot q. \\
\label{4}&Q^\perp\colon & \ot p.\phantom{^\perp} 
          \longrightarrow\ot q.^\perp \,,
\end{eqnarray}
where the first two are related to $\rho$ and the last two 
are related to $\varphi$. 
For example, the parametrization of $\rho$ in terms of $\ot p.^\perp$
means that $\rho$ can be seen as the composition of the following
mappings:
\[
 QP^\perp\ot h.\longrightarrow P^\perp\ot h.\longrightarrow
 Q^\perp P^\perp\ot h. \,.
\]
Later we will also need to consider the mappings
\begin{eqnarray}
\label{5}&P\colon& \ot q.^\perp \longrightarrow  \ot p. \\
\label{6}&P\colon& \ot q. \phantom{^\perp}\longrightarrow  \ot p.   \,.
\end{eqnarray}

Due to the fact that $\ot q.$ and $\ot p.$ are in generic position
the mappings (\ref{1})-(\ref{6}) are bounded, injective and 
their images are dense in the corresponding final subspaces.

We will next give a criterion for the bicontinuity of the mappings
(\ref{1})-(\ref{6}). First note that because $P$ is a basis projection
\[
 \|PQ\|=\|QP\|=\|(\EINS-P)Q\|=\|Q(\EINS-P)\|=:\delta
\] 
and $0< \delta \leq 1$. So we can distinguish between the two
cases: $\delta < 1$ and $\delta=1$.
\begin{pro}\label{Fall<1}
Let $P,Q$ and $\delta$ given as before. If $\delta<1$, then
the mappings (\ref{1})-(\ref{6}) are bicontinuous, in particular
their images coincide with the corresponding final spaces. Moreover,
the relations 
\[
 \|P-Q\|=\|(\EINS-Q)P\|=\|(\EINS-Q)(\EINS-P)\|=\delta
\]
hold.
\end{pro}
\begin{beweis}
This result is a special case of Theorem~6.34 in \cite[p.~56]{bKato95}.
Note that the second alternative stated in Kato's result cannot appear in
the present situation, as a consequence of the fact that $\ot p.$
and $\ot q.$ are in generic position.
\end{beweis}

\begin{rem}
This situation corresponds to the case where the index of $P$ and
$Q$ is 0 (cf.~\cite[Theorem~3.3]{Avron94}).
\end{rem}

\begin{pro}\label{Fall=1}
Let $P,Q$ and $\delta$ given as before. If $\delta=1$, then
the inverse mappings of (\ref{1})-(\ref{6}) are unbounded and densely
defined, i.e.~the images of (\ref{1})-(\ref{6}) are nontrivial 
proper dense sets in the corresponding final subspaces. 
\end{pro}
\begin{beweis}
We will only show the assertion for the mapping (\ref{6}),
since one can easily adapt the following arguments to the other 
cases. Put $A:=QP^\perp Q\hrist\ot q.\in\al L.(\ot q.)$, so that
$A=A^*$ and $A\geq 0$. From 
\[
 \mathrm{spr}\,A=\|A\|=\|QP^\perp P^\perp Q\|=\|P^\perp Q\|^2
 =\delta^2=1
\]
we obtain $1\in\mathrm{spec}\,A$. However, $1$ is not an eigenvalue
of $A$, because $Aq=q$, $q\in\ot q.$, implies
$\mathop{{\rm s-}\lim}\limits_{n\to \infty}\,(QP)^n q=q$
and this means $q\in\ot q.\cap \ot p.^\perp=\{0\}$. Thus
$\mathrm{ker}\,(\EINS_\ot q. -A)=\{ 0\}$ or 
$(\EINS_\ot q. -A)^{-1}$ exists and is unbounded since
$1\notin \mathrm{res}\,A$. Therefore $\vartheta:=
\mathrm{dom}\,(\EINS_\ot q. -A)^{-1}$ is a proper dense subset
in $\ot q.$ and this means $\mathrm{ima}\,(\EINS_\ot q. -A)
=\vartheta=\mathrm{ima}\,(Q -QP^\perp Q)=\mathrm{ima}\,(QPQ)$.
Finally, from the polar decomposition of $PQ$,
\[
 PQ=\mathrm{sgn}\,(PQ)\cdot(QPQ)^{\frac12}\,,
\]
we have that $\mathrm{sgn}\,(PQ)$ maps $\mathrm{ima}\,(QPQ)^{\frac12}$
isometrically onto $\mathrm{ima}\,(PQ)=P\ot q.$. Thus $P\ot q.$ is
a proper dense set in $\ot p.$, i.e.~$P\colon \ot q. \to \ot p.$ is
unbounded invertible.
\end{beweis}

\begin{rem}
(i) Note that if $\mathrm{dim}\,\ot h.< \infty$, then the case 
$\|PQ\|=1$ is not possible, since the corresponding operators
can not have continuous spectrum. It is easy to show that in this
case $\ot p.\cap\ot q.=\{0\}$ iff $\|PQ\|<1$.

(ii) Note also that $\|(\EINS-P)Q\|=1$ implies $\|(\EINS-Q)P\|=1$,
because otherwise by Proposition~\ref{Fall<1} $\|(\EINS-Q)P\|<1$ 
implies $\|(\EINS-P)Q\|<1$.
\end{rem}

\begin{pro}\label{Pro.4.7}
 Let the projections $P,Q$ and the mappings $\varphi,\rho$
be given as before. Then $\varphi,\rho\colon\ \ot q.\to \ot q.^\perp$
are bicontinuous iff $\|PQ\|<1$.
\end{pro}
\begin{beweis}
Suppose that $\|PQ\|<1$, so that by Proposition~\ref{Fall<1} we have
that the mappings (\ref{1})-(\ref{6}) are bicontinuous. But as 
mentioned before we know that $\varphi$ as well as $\rho$
can be seen as composition of the mappings
\begin{eqnarray*}
&\varphi\colon& 
 QP\ot h.\phantom{^\perp}\longrightarrow P\ot h.\phantom{^\perp}
 \longrightarrow Q^\perp P\ot h.\phantom{^\perp} \\
&\rho\colon& 
 QP^\perp\ot h.\longrightarrow P^\perp\ot h.\longrightarrow
 Q^\perp P^\perp\ot h. \,,
\end{eqnarray*}
hence they must be bicontinuous.

In the case that $\rho$ and $\varphi$ are bicontinuous, then
$\mathrm{dom}\,\rho=\mathrm{dom}\,\varphi=\ot q.$ and
$\mathrm{ima}\,\rho=\mathrm{ima}\,\varphi=\ot q.^\perp$.
Finally, Proposition~\ref{Fall=1} implies $\|PQ\|<1$.
\end{beweis}

Motivated by Lemma~\ref{Lem.3.3} we will analyze next the antilinear
mappings defined by the following graphs:
\begin{eqnarray*}
 {\rm gra}\,\beta  &:=& \Big\{(Pq,P\Gamma q)\in\ot p.\times \ot p.
                        \mid q\in\ot q.\Big\} \\
 {\rm gra}\,\alpha &:=& \Big\{(Pq^\perp,-P \Gamma q^\perp) 
                        \in\ot p.\times \ot p.
                        \mid q^\perp\in\ot q.^\perp\Big\}\,.
\end{eqnarray*}
(Note that the r.h.s.~of the preceding equations define indeed 
graphs of antilinear mappings, because the assignments
$q\to Pq$ and $q^\perp\to Pq^\perp$ are injective.)

\begin{lem}\label{Alpha}
The mappings $\alpha,\beta$ defined by the preceding graphs are
anti-linear, injective and closed with dense domains and images
$\mathrm{dom}\,\alpha=\mathrm{ima}\,\alpha=P(\ot q.^\perp)$,
$\mathrm{dom}\,\beta=\mathrm{ima}\,\beta=P\ot q.$.
Further, we have $\alpha^2=\mathrm{id}$, $\beta^2=\mathrm{id}$
on $P(\ot q.^\perp)$ resp.~$P\ot q.$ and $\alpha=\beta^*$.
\end{lem}
\begin{beweis}
We will only prove the last equation, because the other statements
follow immediately from the definition. Now by definition we have
$(p_0,p_1)\in \mathrm{gra}\,\beta^*$ iff
$\langle p_0,P\Gamma q\rangle=\langle Pq, p_1\rangle$
for all $q\in\ot q.$ iff $q^\perp:= p_0 - \Gamma p_1 \in\ot q.^\perp$
iff $(p_0,p_1)\in\mathrm{gra}\,\alpha$.
\end{beweis}

\begin{rem}\label{inclusion}
(i) Recall Subsection~\ref{ModCAR} and denote by $S$
the Tomita operator associated to $(\al M.(\ot q.),\Omega)$. Then
from the preceding result we have
$S\hrist\ot p.\supseteq \beta$ and $S^*\hrist\ot p.\supseteq \alpha$.

(ii) Using the mappings (\ref{5}) and (\ref{6}) we can now state 
similarly as in Proposition~\ref{Pro.4.7} a criterion for the 
bicontinuity of $\alpha,\beta$: the mappings 
$\alpha,\beta$ are bicontinuous iff $\|PQ\|<1$.
\end{rem}

We introduce next the notation
\[
 \Delta_\ot p.:=\beta^*\!\beta\,,
\]
since it will later turn out that $\Delta_\ot p.$ is actually 
the modular operator restricted to the one-particle Hilbert space
$\ot p.$.

\begin{teo}\label{beta*beta}
The mapping $\Delta_\ot p.\colon\ \ot p.\to \ot p.$ is a densely defined
linear positive self-adjoint operator on $\ot p.$ with graph
\[
 {\rm gra}\,\Delta_\ot p. = \Big\{(PQp,PQ^\perp p) \mid p\in\ot p.\Big\}\,.
\]
Moreover, $\Delta_\ot p.^{-1}=\beta\beta^*=\alpha^*\!\alpha$.
\end{teo}
\begin{beweis}
We will compute first the domain of $\beta^*\!\beta$. Recalling
that $\beta^* =\alpha$ we have
\begin{eqnarray*}
\mathrm{dom}\,(\Delta_\ot p.) 
             &=&\Big\{Pq\mid q\in\ot q.\;\mathrm{and}\;
                 P\Gamma q\in\mathrm{dom}\,\alpha=P(\ot q.^\perp)\Big\}\\[2mm]
              &=&\Big\{Pq\mid q\in\ot q.\;\mathrm{and}\;
   P\Gamma q=Pq^\perp\;\mathrm{for~some}\;q^\perp\in\ot q.^\perp\Big\}\\[2mm]
             &=&\Big\{Pq\mid q\in\ot q.\;\mathrm{and}\;
                 \Gamma q\in\mathrm{dom}\,\rho=Q(\ot p.^\perp)\Big\}\\[2mm]
             &=&\Big\{Pq\mid q\in\ot q.\;\mathrm{and}\;
                 q\in \Gamma Q(\ot p.^\perp)=Q(\Gamma \ot p.^\perp)
                 =Q\ot p.\Big\}\\[2mm]
             &=&PQ\ot p.\kern3mm= \kern3mmPQP\ot h.\,,
\end{eqnarray*}
which is dense in $\ot p.$. Furthermore, since $P\Gamma Q p=-PQ^\perp
\Gamma p$, $p\in\ot p.$ (recall $P\Gamma p=0$, $p\in\ot p.$), we
have
\[
 \Delta_\ot p.(PQp)=\alpha\Big(P\Gamma Qp\Big)
 =-\alpha\Big(P Q^\perp \Gamma p\Big)=P\Gamma Q^\perp \Gamma p
 =PQ^\perp p\,,\quad p\in\ot p.\,.
\] 
The last equations concerning the inverse of $\Delta_\ot p.$
follow from the preceding computation and from the fact that
$\alpha^2=\mathrm{id}$ and $\beta^2=\mathrm{id}$ on the corresponding
domains (recall Lemma~\ref{Alpha}).
\end{beweis}

Note that $\mathrm{dom}\,\Delta_\ot p.=PQ\ot p.$ is dense in $\ot p.$
and that $\Delta_\ot p.^{-\frac12}=|\alpha|$ hence
$\mathrm{dom}\,\Delta_\ot p.^{-\frac12}=P(\ot q.^\perp)$.
Next we will calculate the graph of the positive self-adjoint operator
$\varphi^*\varphi\colon\ot q.\to \ot q.$.
\begin{pro}\label{Provarphi*varphi}
The graph of $\varphi^*\varphi$ is given by 
\[
{\rm gra}\,\varphi^*\!\varphi 
    = \Big\{(QPq,QP^\perp q) \mid q\in\ot q.\Big\}\,.
\]
\end{pro}
\begin{beweis}
We begin computing $\mathrm{dom}\,(\varphi^*\!\varphi)$. Since by 
Lemma~\ref{Rho-1} $\varphi^*=\rho^{-1}$ we have
\begin{eqnarray*}
\mathrm{dom}\,(\varphi^*\!\varphi) 
            &=&\Big\{Qp\mid p\in\ot p.\;\mathrm{and}\;
      Q^\perp p\in\mathrm{dom}\,\rho^{-1}=Q^\perp\ot p.^\perp\Big\}\\[2mm]
            &=&\Big\{Qp\mid p\in\ot p.\;\mathrm{and}\;
      Q^\perp p= Q^\perp p^\perp
      \;\mathrm{for~some}\;p^\perp\in\ot p.^\perp\Big\}\\[2mm]
          &=&\Big\{Qp\mid p= Pq \;\mathrm{for~some}\;q\in\ot q.\Big\}\\[2mm]
          &=&QP\ot q.\kern3mm= \kern3mm QPQ\ot h.\,,
\end{eqnarray*}
where for the third equation we have used Corollary~\ref{Lambda}.
Using again this corollary we can calculate
\[
 \rho^{-1}\Big(\varphi(QPq)\Big)=\rho^{-1}(Q^\perp Pq)
 =\rho^{-1}\Big( Q^\perp (-P^\perp q)\Big)=QP^\perp q\,,
  \quad q\in\ot q.\, ,
\]
and the proof is concluded.
\end{beweis}

Now we can relate $\varphi$ and $\varphi^*\varphi$ with $\Delta_\ot p.$
just computing the orthogonal decomposition of $\varphi(q)$, 
$q\in\mathrm{dom}\,\varphi$, resp.~$\varphi^*\varphi(q)$, 
$q\in\mathrm{dom}\,\varphi^*\varphi$, 
w.r.t.~$\ot h.=\ot p.\oplus\ot p.^\perp$.
\begin{cor}\label{Deltap}
Using the notation before we have the following formulas:
\begin{eqnarray}
\label{varphi*varphi} \varphi(Qp)
  &=& \Delta_\ot p.(PQp)-P^\perp Qp\,,\quad p\in\ot p.\,,  \\[2mm]
\nonumber   \Big(\varphi^*\varphi\Big)(QPq)
  &=& \Delta_\ot p.(PQPq)+\Gamma\,\Delta_\ot p.^{-1}(P\Gamma QPq)
      \,,\quad q\in\ot q.\,.
\end{eqnarray}
\end{cor}
\begin{beweis}
From Lemma~\ref{Rho-1} as well as Theorem~\ref{beta*beta} we have
\[
 \varphi (Qp)=Q^\perp p= PQ^\perp p+ P^\perp Q^\perp p
  =\Delta_\ot p.(PQp)-P^\perp Qp\,,\quad p\in\ot p.\,. 
\]
Further, from the preceding proposition we also have for any $q\in\ot q.$
\begin{eqnarray*}
\Big(\varphi^*\varphi\Big)(QPq)
  &=& QP^\perp q \kern3mm =\kern3mm PQP^\perp q+P^\perp QP^\perp q \\
  &=& PQ^\perp P q + \Gamma PQP \Gamma q \\
  &=& \Delta_\ot p.(PQPq)+\Gamma\,\Delta_\ot p.^{-1}(PQ^\perp P\Gamma q)\\
  &=& \Delta_\ot p.(PQPq)+\Gamma\,\Delta_\ot p.^{-1}(PQP^\perp\Gamma q)\\
  &=& \Delta_\ot p.(PQPq)+\Gamma\,\Delta_\ot p.^{-1}(P\Gamma QPq)\,,
\end{eqnarray*}
which proves the second formula.
\end{beweis}

We will give next two formulas in terms of $\Delta_\ot p.^\frac12$
for the components of the polar decomposition of $\varphi$. 
Denote $\varphi=\mathrm{sgn}\,\varphi\cdot |\varphi|$, where as usual
$|\varphi|:=(\varphi^*\varphi)^\frac12$. Recall from the results in 
this section that
\begin{eqnarray*}
\mathrm{dom}\,\varphi\kern2mm=\kern2mm
\mathrm{dom}\,|\varphi|\kern2mm=\kern2mm Q\ot p.
&\supseteq& QP\ot q.\kern2mm=\kern2mm\mathrm{dom}\,(\varphi^*\varphi) \\[2mm]
\mathrm{dom}\,\beta\kern2mm=\kern2mm
\mathrm{dom}\,|\beta|\kern2mm=\kern2mm P\ot q.
&\supseteq& PQ\ot p.\kern2mm=\kern2mm \mathrm{dom}\,\Delta_\ot p.\,.
\end{eqnarray*} 

\begin{teo}\label{Teo.4.15}
With the notation above we have:
\begin{eqnarray}
\label{|varphi|} |\varphi|(q)
  &=& \Delta_\ot p.^\frac12(Pq)+\Gamma\,\Delta_\ot p.^{-\frac12}(P\Gamma q)
      \,,\quad q\in\mathrm{dom}\,\varphi^*\varphi = QP\ot q.\,,  \\[2mm]
\label{sgnvarphi} \mathrm{sgn}\,\varphi(q)
  &=& \Delta_\ot p.^\frac12(Pq)-\Gamma\,\Delta_\ot p.^{\frac12}(P\Gamma q)
      \,,\quad q\in\ot q.\,.
\end{eqnarray}
Moreover $\mathrm{sgn}\,\varphi$ is an isometry of $\ot q.$ onto
$\ot q.^\perp$, i.e.~$(\mathrm{sgn}\,\varphi)^*\mathrm{sgn}\,\varphi=Q$
and $\mathrm{sgn}\,\varphi(\mathrm{sgn}\,\varphi)^*=Q^\perp$.
\end{teo}
\begin{beweis}
From the explicit knowledge of all the domains of the mappings used
before it is easily seen that the formulas are well-defined. Further,
recall Proposition~\ref{Provarphi*varphi} and Corollary~\ref{Deltap} and 
the fact that the two terms of the r.h.s.~of the above formulas 
correspond to the decomposition of $\ot h.$ in terms of $P\ot h.$
and $P^\perp\ot h.$. Applying now for $q\in\mathrm{dom}\,\varphi^*\varphi$
twice the r.h.s.~of (\ref{|varphi|}) we get
\begin{eqnarray*}
 \lefteqn{\kern-3.5cm
 \Delta_\ot p.^\frac12 P \Big(
 \Delta_\ot p.^\frac12(Pq)+\Gamma\,\Delta_\ot p.^{-\frac12}(P\Gamma q)
 \Big)+\Gamma\,\Delta_\ot p.^{-\frac12} P\Gamma \Big(
 \Delta_\ot p.^\frac12(Pq)+\Gamma\,\Delta_\ot p.^{-\frac12}(P\Gamma q)\Big)}\\
   &=& \Delta_\ot p. (Pq)+\Gamma \Delta_\ot p.^{-1}(P\Gamma q)
     \kern2mm=\kern2mm \Big(\varphi^*\varphi\Big)(q)\,,
\end{eqnarray*}
which shows the first formula. To prove the second one note first
that for $q'=|\varphi|(q)$, $q\in \mathrm{dom}\,\varphi^*\varphi$, and
using Eq.~(\ref{|varphi|}) as well as Corollary~\ref{Deltap}
we obtain from a similar calculation as before that
\[
\Big(\Delta_\ot p.^\frac12 P-\Gamma\,\Delta_\ot p.^{\frac12} P\Gamma\Big) 
|\varphi|(q')=\Big(\Delta_\ot p. P- P^\perp\Big)(q')=\varphi(q')\,.
\]
Thus the r.h.s.~and the l.h.s.~of (\ref{sgnvarphi}) coincide on the dense
subspace $\mathrm{ima}\,\varphi^*\varphi$. 
Finally, the fact that the r.h.s.~is also well defined for all 
$q\in\ot q.$ (recall that $\mathrm{dom}\,\Delta_\ot p.^\frac12
=\mathrm{dom}\,\beta=P\ot q.$) and that
$\mathrm{sgn}\,\varphi$ maps isometrically the dense
subspace $\mathrm{ima}\,|\varphi |\subseteq \ot q.$ onto the
dense subspace $\mathrm{ima}\,\varphi =Q^\perp \ot p.\subseteq 
\ot q.^\perp$ proves formula (\ref{sgnvarphi}). Therefore
$Q$ is the initial projection of $\mathrm{sgn}\,\varphi$ 
and $Q^\perp$ is the corresponding final projection.
\end{beweis}

\begin{rem}
Recall that $\mathrm{dom}\,\varphi^*\varphi$ is a core for $|\varphi|$
and note that the r.h.s.~of formula (\ref{|varphi|}) {\em can not}
be extended to the whole $\mathrm{dom}\,|\varphi|$.
\end{rem}

Finally, we consider the mapping 
\begin{equation}\label{W}
 \ot q.\ni q\mapsto Wq:=(\EINS+\Delta_\ot p.)^\frac12\,Pq \in\ot p.\,.
\end{equation}

\begin{lem}
$W$ is an isometry from $\ot q.$ onto $\ot p.$.
\end{lem}
\begin{beweis}
First choose $q_1,q_2\in Q\ot p.$ which is dense in $\ot q.$. Now using
\[
\langle Pq_1,\Delta_\ot p.(Pq_2)\rangle
    =\langle Pq_1,\beta^*\!\beta(Pq_2)\rangle
    =\langle \beta (Pq_2),\beta(Pq_1)\rangle
    =\langle P^\perp q_1, P^\perp q_2\rangle
\] 
we obtain
\[
\Big\langle(\EINS+\Delta_\ot p.)^\frac12 Pq_1,
           (\EINS+\Delta_\ot p.)^\frac12 Pq_2\Big\rangle
    = \langle Pq_1,(\EINS+\Delta_\ot p.) Pq_2\rangle
     = \langle q_1, q_2\rangle\,, q_1,q_2\in Q\ot p.\,.
\] 
Further, for any $p\in\ot p.$ we have from Theorem~\ref{beta*beta}
that $(\EINS+\Delta_\ot p.)(PQp)=PQp+PQ^\perp p=p$. This implies that
$\mathrm{ima}\,(\EINS+\Delta_\ot p.)^\frac12 =\ot p.$, since
\[
\ot p.=\mathrm{ima}\,(\EINS+\Delta_\ot p.)\subseteq
       \mathrm{ima}\,(\EINS+\Delta_\ot p.)^\frac12\subseteq \ot p.\,. 
\]
Therefore (\ref{W}) is the isometric extension of $W\hrist Q\ot p.$.
\end{beweis}

Using now the isometry $W$ we conclude this section showing the unitary
equivalence of $|\varphi|$ and $\Delta_\ot p.^\frac12$.

\begin{teo}\label{Equivalence}
With the preceding notation we have
\[
 W\,|\varphi|(q)=\Delta_\ot p.^\frac12\,W(q)\,,
                 \quad q\in\mathrm{dom}\,|\varphi|\,.
\]
\end{teo}
\begin{beweis}
For any $q\in\mathrm{dom}\,\varphi^*\varphi$, which is a core of 
$|\varphi|$, we may use (\ref{|varphi|}) and in this case
\[
 W\,|\varphi|(q)=(\EINS+\Delta_\ot p.)^\frac12 P
    \Big(\Delta_\ot p.^\frac12(Pq)+
         \Gamma\,\Delta_\ot p.^{-\frac12}(P\Gamma q)\Big)
     =(\EINS+\Delta_\ot p.)^\frac12 \,\Delta_\ot p.^\frac12(Pq)
     = \Delta_\ot p.^\frac12 W(q)\,.
\]
But this implies that $W |\varphi| W^*\subseteq \Delta_\ot p.^\frac12$
and since the l.h.s.~as well as the r.h.s.~of the preceding inclusion
are self-adjoint operators, we must actually have the equality
$W |\varphi| W^*= \Delta_\ot p.^\frac12$.
\end{beweis}

%%%%%%%%%%%%%%%%%%%%%%%%%%%%%%%%%%%%%%%%%%%%%%%%%%%%%%%%%%%%%%%%%%%%%%%%%%%%%%
%%%%%%%%%%%%%%%%%%%%%%%%%%%%%%%%%%%%%%%%%%%%%%%%%%%%%%%%%%%%%%%%%%%%%%%%%%%%%%
\section{Twisted duality. The generic position case}\label{AbsDua}

We begin the proof of twisted duality considering first one of the
extremal cases that may appear in the Halmos decomposition (\ref{Halmos}).
For this assume that $(\ot h.,\Gamma)$, $P$ and
$Q$ are given as in the preceding section. In particular $\ot p.$
and $\ot q.$ are in generic position, so that 
by Proposition~\ref{Teo.3.2} the modular theory is
well defined for the pair $(\al M.(\ot q.),\Omega)$. Denote as usual
by $S=J\Delta^\frac12$ the polar decomposition of the Tomita operator.

We prove first that the different modular objects leave the 
$n$-particle submanifolds $\mathop{\land}\limits^n(P\ot q.)$ invariant. 
This fact is well known in the context of CCR-algebras 
\cite{pLeyland78}, where one can use the so-called 
exponential vectors which are specially well-adapted to the 
Weyl operators.

\begin{pro}\label{NTeilchen}
Let $q_1,\ldots,q_n\in \ot q.$ and $q^\perp_1,\ldots,
q^\perp_n\in \ot q.^\perp$. Then the following equations hold
\begin{eqnarray*}
S(P q_1\land \ldots \land Pq_n)
    &=& P \Gamma q_n\land \ldots\land P\Gamma q_1\kern2mm=\kern2mm
        S(P q_n)\land \ldots \land S(Pq_1 ) \\
S^*(P q_1^\perp\land \ldots \land Pq_n^\perp)
    &=& S^*(P q_n^\perp)\land \ldots \land S^*(Pq_1^\perp)
\end{eqnarray*}
\end{pro}
\begin{beweis}
The proof is done by induction on the number of vectors in the wedge 
product. For $n=1$ the above equations are trivially satisfied 
(cf.~Lemma~\ref{Lem.3.3}). We will now concentrate on the first
formula since one can argue similarly for $S^*$. Suppose that the
first expression holds for a number of vectors 
$\leq n-1$. Then applying this induction
hypothesis as well as Proposition~\ref{Formel} we get
\begin{eqnarray}
\nonumber\lefteqn{S(P \Gamma q_n\land \ldots\land P\Gamma q_1)}\\[3mm]
\nonumber  &=& S\Bigg(a(q_n)\cdot\ldots\cdot a(q_1)\,\Omega \;-
               \sum\limits_{\mbox{\tiny $\begin{array}{c}\pi\in\ot S._{n,p}
               \\[1mm] p\geq 1\end{array}$}}
                   \!\!\!\!({\rm sgn}\,\pi)\;
  \prod\limits_{l=1}^p \;\langle Pq_{\alpha_l}\,,\,P\Gamma q_{\beta_l}\rangle
  \, P\Gamma q_{j_1} \land \ldots \land P\Gamma q_{j_k}
               \Bigg)  \\
\label{Big-}&=& a(\Gamma q_1)\cdot\ldots\cdot a(\Gamma q_n)\,\Omega \;-
       \sum\limits_{\mbox{\tiny $\begin{array}{c}\pi\in\ot S._{n,p}
               \\[1mm] p\geq 1\end{array}$}}
               \!\!\!\!({\rm sgn}\,\pi)\;
  \prod\limits_{l=1}^p \;\langle P\Gamma q_{\beta_l}\,,\,Pq_{\alpha_l}\rangle
  \, P q_{j_k} \land \ldots \land P q_{j_1}
\end{eqnarray}
and recall that the indices specified by $\pi\in\ot S._{n,\,p}$
satisfy $\alpha_1>\ldots >\alpha_p$, $\alpha_l>\beta_l$, $l=1,\ldots,
p$ and $n\geq j_1>j_2>\ldots >j_k\geq 1$. But we can now apply again
Proposition~\ref{Formel} to the first term of the preceding sum
and we get
\begin{equation}\label{insert}
a(\Gamma q_1)\cdot\ldots\cdot a(\Gamma q_n)\,\Omega 
 = P q_1\land \ldots \land Pq_n +
   \sum\limits_{\mbox{\tiny $\begin{array}{c}\pi'
               \\[1mm] p\geq 1\end{array}$}}
               \!\!\!\!({\rm sgn}\,\pi')\;
  \prod\limits_{l=1}^p \;\langle P\Gamma q_{\alpha_l'}\,,\,
  Pq_{\beta_l'}\rangle\, P q_{j_1'} \land \ldots \land P q_{j_k'}\,,
\end{equation}
where now $\pi'=\mbox{\scriptsize
       $\left(\kern-1.5mm {\begin{array}{cccccccc}
       1 & 2       &\cdots & 2p-1      & 2p &  n-k+1   & \cdots & n \\
\alpha_1' &\beta_1'&\cdots & \alpha_p' &\beta_p' & j_1' & \ldots & j_k'   
          \end{array}}\kern-1.5mm\right)$}$
with $\alpha_l'<\beta_l'$, $l=1,\ldots,p$, 
$1\leq j_1'<j_2'<\ldots <j_k'\leq n$ and we may reorganize the 
scalar products such that $\beta_1'<\ldots < \beta_p'$. We can next
associate bijectively these permutations with elements in
$\ot S._{n,\,p}$ by means of 
\[
 \pi' \mapsto \pi_0 =
 \mbox{\scriptsize
       $\left(\kern-1.5mm {\begin{array}{cccccccc}
       n & n-1     &\cdots & n-2p+2  & n-2p+1 & k   & \cdots & 1 \\
\beta_p' &\alpha_p' &\cdots &\beta_1' &\alpha_1' & j_k' & \ldots & j_1'   
          \end{array}}\kern-1.5mm\right)$} \in \ot S._{n,\,p}\,.
\]
Since $\mathrm{sgn}\,\pi'=\mathrm{sgn}\,\pi_0'$ we have inserting
(\ref{insert}) in (\ref{Big-}) that all vectors with particle number
less than $n$  cancel so that
\[
S(P \Gamma q_n\land \ldots\land P\Gamma q_1) =
    P q_1\land \ldots \land Pq_n
\]
and the proof is concluded.
\end{beweis}

Recall from Remark~\ref{inclusion}~(i)
$S\hrist\ot p.\supseteq \beta$ and $S^*\hrist\ot p.\supseteq \alpha$.
We will show next that actually the equality holds. 
Let $P_n$ denote the projection of $\ot F.$ onto the $n$-particle
subspace $\mathop{\land}\limits^n \ot p.$. Then the family of
orthoprojections $\{P_n\}_{n\in\N}$ is mutually orthogonal and
$\sum_{n=0}^\infty P_n =\EINS_{\ot F.}$. Further we define
the operator $\Sfin$ by 
\begin{eqnarray*}
 \mathrm{dom}\,\Sfin &:=&\mathrm{span}\,
    \Big\{ a(q_1)\cdot\ldots\cdot a(q_n)\Omega\mid
    q_1,\ldots,q_n\in\ot q.\;,\;n\in\N\cup\{0\}\Big\}\\
   \Sfin x &:=& Sx\,,\quad x\in\mathrm{dom}\,\Sfin\,.
\end{eqnarray*}
\begin{lem}
$\mathrm{dom}\,\Sfin$ is a core for the Tomita operator $S$.
\end{lem}
\begin{beweis}
Put $\al C.:=\mathrm{dom}\,\Sfin$ and denote by $\al A.(\ot q.)$
the *-algebra generated by $\{a(q)\mid q\in\ot q.\}$, so that
$\al M.(\ot q.)=\al A.(\ot q.)''$. Further recall that if 
$S_0(M\Omega):=M^*\Omega$, $M\in\al M.(\ot q.)$, then the graph
of the Tomita operator $S=\mathrm{clo}\,S_0$ can be written as
\[
 \mathrm{gra}\,S=\mathrm{clo}_w(\mathrm{gra}\,S_0)=\mathrm{clo}_w
 \Big\{(M\Omega,M^*\Omega)\mid M\in \al M.(\ot q.)\Big\}\,,
\]
where $\mathrm{clo}_w$ denotes the closure in the weak operator topology.
Thus to prove the core property of $\al C.$, 
i.e.~$\mathrm{clo}\,(S\hrist\al C.)=S$, we need to show that
\[
 \mathrm{clo}_w\Big\{(M\Omega,M^*\Omega)\mid M\in\al M.(\ot q.)\Big\}
 =\mathrm{clo}_w\Big\{(A\Omega,A^*\Omega)\mid A\in\al A.(\ot q.)\Big\}\,.
\]

Now for each $M\in\al M.(\ot q.)$ there exists a sequence
$\{A_n\}_{n\in\N}\subset\al A.(\ot q.)$ with 
$|\langle x,(M-A_n)y\rangle|<\epsilon$ for all $x,y\in\ot F.$
if $n$ is sufficiently large.
This implies that for all $(x,y)\in\ot F.\times\ot F.$, we have
\[
 \Big|\Big\langle (M\Omega,M^*\Omega),(x,y) \Big\rangle -
 \Big\langle (A_n\Omega,A_n^*\Omega),(x,y) \Big\rangle\Big|
 = \Big|\Big\langle (M-A_n)\Omega,x \Big\rangle -
  \Big\langle (M^*-A_n^*)\Omega,y \Big\rangle\Big|
 \leq 2\epsilon \,,
\]
where we have used that in the weak operator topology
if $A_n\to M$, then $A_n^*\to M^*$. Therefore if 
$(M_l\Omega,M_l^*\Omega)\stackrel{l}{\longrightarrow} (x_0,y_0)$
weakly, then for each $l$ there exists a sequence 
$\{A_{(n,l)}\}_{n\in \N}\subset\al A.(\ot q.)$ such that 
$(A_{(n,l)}\Omega,A_{(n,l)}^*\Omega)\stackrel{n}{\longrightarrow} 
(M_l\Omega,M_l^*\Omega)$ weakly. Finally, the estimate
\begin{eqnarray*}
 \lefteqn{\!\!\!\!\!\!\!\!\!\!\!\!\!\!\!\!\!\!\!\!\!\!\!\!\!\!\!\!\!
  \Big|\Big\langle (A_{(n,l)}\Omega,A_{(n,l)}^*\Omega),
  (x,y)\Big\rangle -\Big\langle (x_0,y_0),(x,y) \Big\rangle\Big|} \\
   &\leq& \Big|\Big\langle (A_{(n,l)}\Omega,A_{(n,l)}^*\Omega),
         (x,y)\Big\rangle -\Big\langle (M_l\Omega,M_l^*\Omega),
         (x,y)\Big\rangle\Big| \\
   &&\!\!\!\!\!+\Big|\Big\langle (M_l\Omega,M_l^*\Omega),(x,y)\Big\rangle - 
         \Big\langle (x_0,y_0),(x,y) \Big\rangle\Big|
\end{eqnarray*}
completes the proof.
\end{beweis}

From Propositions~\ref{NTeilchen} and \ref{Formel} we obtain further
\begin{eqnarray*}
P_n\,\mathrm{dom}\,\Sfin 
     &\subseteq& \mathrm{dom}\,\Sfin\,,\quad n=0,1,\ldots\,,  \\
P_n\,\mathrm{dom}\,\Sfin 
     &\subset& P_n\,\ot F.\;\;\mathrm{is~dense}\,,\quad n=0,1,\ldots\,,  \\
P_n\,\Sfin \, P_n  &=& \Sfin\,P_n\,,\quad n=0,1,\ldots\,.
\end{eqnarray*}
Note that for $x\in\mathrm{dom}\,\Sfin$ the series
$x=\sum_{n=0}^\infty P_n x$ is a finite sum. Further we consider
the operators $\Sfin(n):P_n\ot F.\to P_n\ot F.$ by
$\mathrm{dom}\,\Sfin(n):= P_n\,\mathrm{dom}\,\Sfin$ and
$\Sfin(n) x :=\Sfin x\,,x\in\mathrm{dom}\,\Sfin (n)$.
Recall from Lemma~\ref{Alpha} that $\Sfin (1)=\beta$ is a 
closed densely defined operator from $\ot p.=P_1\ot F.$ 
into $\ot p.$ and by the preceding arguments we 
have
\begin{equation}\label{ADS}
 \Sfin =\mathop{\oplus}\limits_{n=0}^\infty\Sfin(n)\qquad
 \mathrm{(algebraic~direct~sum)}\,.
\end{equation}
We can now state:
\begin{pro}\label{5.3}
The Tomita operator $S$ and its adjoint $S^*$ 
can be restricted to the $n$-particles subspaces $P_n\ot F.$. 
We have 
\[
  S\hrist P_n\ot F.=\mathrm{clo}\,\Sfin(n)\quad\mathrm{and}\quad
  S^*\hrist P_n\ot F.=\mathrm{clo}\,\Sfin^*(n)\,,\quad n=0,1,\ldots\,.
\]
In particular $S\hrist\ot p.=\beta$ and $S^*\hrist \ot p.=\alpha$.
\end{pro}
\begin{beweis}
From (\ref{ADS}) we obtain immediately that 
\begin{equation}\label{HS}
\mathrm{gra}\,S=\mathop{\oplus}\limits_{n=0}^\infty
                 \mathrm{clo}\,\mathrm{gra}\,\Sfin(n)
               =\C \Omega\oplus\mathrm{gra}\,\Sfin(1)\oplus
                \sum\limits_{n=2}^\infty\mathrm{clo}\,\mathrm{gra}\,\Sfin(n)
                \qquad\mathrm{(Hilbert~sum)}\,.
\end{equation}
Note that $\mathrm{clo}\,\mathrm{gra}\,\Sfin(n)=
\mathrm{gra}\,\mathrm{clo}\,\Sfin(n)$, $n\geq 2$ and that 
now the direct sum (\ref{HS}), in contrast to (\ref{ADS}),
is the Hilbert sum of these subspaces. Eq.~(\ref{HS}) implies
immediately the assertion that $S$ can be restricted to $P_n\ot F.$
and that the restriction coincides with $\mathrm{clo}\,\Sfin(n)$, in
particular $S\hrist\ot p.=\beta$. The statements concerning $S^*$
are shown analogously.
\end{beweis}

\begin{cor}\label{CorDJ}
Let $S=J\Delta^\frac12$ be the polar decomposition of the Tomita operator.
The modular operator $\Delta =S^*S$ and the modular conjugation $J$ 
can be restricted to the respective $n$-particle subspaces. 
In particular we have:
\begin{itemize}
\item[(i)] Modular operator: $\Delta\hrist\ot p.=\Delta_\ot p.$ 
  (recall from the preceding section that $\Delta_\ot p.=\beta^*\!\beta$),   
  $\mathrm{dom}\,\Delta\hrist P_n\ot F.=\mathop{\land}\limits^n
  \mathrm{dom}\,\Delta_\ot p.$ and 
 \begin{equation}\label{MOn}
  \Delta (p_1\land\ldots\land p_n)=(\Delta_\ot p.\,p_1)
  \land\ldots\land(\Delta_\ot p.\,p_n)\,,\quad 
  p_1,\ldots,p_n\in\mathrm{dom}\,\Delta_\ot p.=PQ\ot p.\,.
 \end{equation}
\item[(ii)] Modular conjugation:
\begin{equation}\label{MCn}
J (p_1\land\ldots\land p_n)=(Jp_n)
  \land\ldots\land (Jp_1)\,,\quad p_1,\ldots,p_n\in\ot p.\,.
\end{equation}
\end{itemize}
\end{cor}
\begin{beweis}
Note first that from Proposition~\ref{5.3} and Lemma~\ref{Alpha}
we have $S^*S\hrist\ot p.=\alpha\beta=\beta^*\!\beta$. Further Eq.~(\ref{MOn}) 
follows from Propositions~\ref{NTeilchen} and \ref{5.3}.
Next, applying formula (\ref{MCn}) to the $n$-particle vectors
$\Delta^\frac12 (p_1\land\ldots\land p_n)$, where $p_k:=
\Delta_\ot p.^\frac12\,p_k'$, $p_k'\in\mathrm{dom}\,\Delta_\ot p.$,
$k=1,\ldots,n$, we obtain $S(p_1\land\ldots\land p_n)$.
Thus Eq.~(\ref{MCn}) coincides with 
$J\hrist P_n\ot F.$ on the dense set $(\mathrm{ima}\,\Delta_\ot p.)
\land\ldots\land (\mathrm{ima}\,\Delta_\ot p.)$, which concludes the proof.
\end{beweis}

For the next definition recall the formulas concerning 
$\mathrm{sgn}\,\varphi$ in Theorem~\ref{Teo.4.15}.              
\begin{defi}\label{V}
Define the following anti-linear isometry from $\ot q.$ onto 
$\ot q.^\perp$:  
\[
 Vq:=-i(\Gamma\,\mathrm{sgn}\,\varphi)(q)
    =i \Big(\Delta_\ot p.^\frac12 P\Gamma q
        -\Gamma\,\Delta_\ot p.^{\frac12}P q\Big)\,,\quad q\in\ot q.\,.
\]
\end{defi}

\begin{teo}
With the preceding definition we have for any $q\in\ot q.$
\begin{equation}\label{JaZ}
 J \,a(q)\, J = \widetilde{Z}\, a(Vq)\, \widetilde{Z}^*\,.
\end{equation}
\end{teo}
\begin{beweis}
Since $\mathop{\oplus}\limits_{n=0}^\infty\mathop{\land}\limits^n P\ot q.$
(algebraic direct sum) is dense in $\ot F.$ and the operators on 
both sides of Eq.~(\ref{JaZ}) are bounded it is enough to show the
preceding relation for the $n$-particle vectors in 
$\mathop{\land}\limits^n P\ot q.$. For $q_1,\ldots,q_n\in\ot q.$ 
and using Corollary~\ref{CorDJ}~(ii) the l.h.s.~of the equation reads
\begin{eqnarray*}
 \lefteqn{\Big(J \,a(q)\, J \Big)(Pq_1\land\ldots\land Pq_n)} \\
    &=& Pq_1\land\ldots\land Pq_n\land JP\Gamma q
        + \sum\limits_{r=n}^{1}\,(-1)^{n-r}\,\langle JPq,Pq_r\rangle 
        \;Pq_1 \land\ldots\land\widehat{Pq_r} \land \ldots \land Pq_n\,.
\end{eqnarray*}
To compute the r.h.s.~recall the definition of the function $\eta$ in
Subsection~\ref{Q}.
\begin{eqnarray*}
\lefteqn{\Big( \widetilde{Z}\, a(Vq)\, \widetilde{Z}^*\Big)(Pq_1\land\ldots\land Pq_n)}\\
  &=& -i\overline{\eta(n)}\,\widetilde{Z}\cdot\Big(
     -c(\Delta_\ot p.^{\frac12}P q)^*Pq_1\land\ldots\land Pq_n
   +c(\Delta_\ot p.^{\frac12}P\Gamma q) Pq_1\land\ldots\land Pq_n \Big)\\ 
  &=& Pq_1\land\ldots\land Pq_n\land\Delta_\ot p.^{\frac12}P q
     +\sum\limits_{r=1}^{n}\,(-1)^{n-r}\,
      \langle \Delta_\ot p.^{\frac12}P\Gamma q,Pq_r\rangle 
      \;Pq_1 \land\ldots\land\widehat{Pq_r} \land \ldots \land Pq_n\,,
\end{eqnarray*}
where for the last equation we have used that 
$-i\overline{\eta(n)}\eta(n+1)=-i\overline{\eta(n)}\eta(n-1)=(-1)^{n+1}$.
Finally, the equality of both sides follows from the fact that
$\Delta_\ot p.^{\frac12}P q=JSPq=JP\Gamma q$, $q\in\ot q.$.
\end{beweis}

\begin{rem}\label{rel1}
As expected we can relate the mapping $V$ defined before with the
mapping $j$ (the modular conjugation restricted to the one-particle
Hilbert space) used in \cite[p.~738]{Foit83}. 
Indeed, considering the projection onto the one-particle 
Hilbert space we have the relation
\[
 J(Pq)=\Delta_\ot p.^{\frac12}S(Pq)=\Delta_\ot p.^{\frac12}(P\Gamma q)
      =-i PV(q)\,,\quad q\in\ot q.\,.
\]
\end{rem}

\begin{teo} (Twisted Duality) \label{TD0}
Let $\al M.(\ot q.)$ be the von Neumann algebra given at the beginning
of this section. Then
\[
 \al M.(\ot q.)'= \widetilde{Z}\,\al M.(\ot q.^\perp)\, \widetilde{Z}^*\,.
\]
\end{teo}
\begin{beweis}
From the last theorem and using standard results in modular theory
we have
\begin{eqnarray*}
\al M.(\ot q.)' 
  &=& J\, \al M.(\ot q.)\, J 
    \kern2mm =\kern2mm J\, \{ a(q)\mid q\in\ot q.\}''\; J \\[2mm]
  &=& \{ J\,a(q)\,J\mid q\in\ot q.\}'' \\[2mm]
  &=& \{ \widetilde{Z}\,a(Vq)\,\widetilde{Z}^*\mid q\in\ot q.\}''
     \kern2mm =\kern2mm 
     \widetilde{Z}\,\{ a(Vq)\mid q\in\ot q.\}''\; \widetilde{Z}^* \\[2mm]
  &=& \widetilde{Z}\,\al M.(\ot q.^\perp)\, \widetilde{Z}^*\,,
\end{eqnarray*}
where we have used that $V$ is an anti-linear isometry from $\ot q.$
onto $\ot q.^\perp$.
\end{beweis}

\begin{rem}
If one does not want to bother about domain questions, there is 
possibly an alternative way to show the 
preceding result. Indeed, one can 
first prove the statements in this section for finite dimensional
subspaces $\ot q._n$ of $\ot q.$ and then apply the AFD-property
of $\al M.(\ot q.)$ as in the proof of \cite[Theorem~15.1.3]{bBaumgaertel92}.
\end{rem}

We will finally show that the formulas established in the previous
and present sections also apply to the localized algebras that appear
in the context of fermionic free nets (cf.~\cite{Lledo95,Lledo01}).
Let $\al O.\subset\R^4$ be a double cone in Minkowski space
and denote by $\overline{\ot q.(\al O.)}$ the closure of the subspaces 
$\ot q.(\al O.)$ of the reference Hilbert space $(\ot h.,\Gamma)$. 
The subspaces $\ot q.(\al O.)$ are defined 
in terms of the embeddings that characterize
the free nets (essentially Fourier transformation of $C^\infty$ functions
with compact support restricted to the positive mass shell/light cone).
It is easily shown that $\Gamma\ot q.(\al O.)=\ot q.(\al O.)$,
hence $\Gamma\overline{\ot q.(\al O.)}=\overline{\ot q.(\al O.)}$.
Further the localized C*-algebras are again CAR-algebras, i.e.
\[
 \al A.(\al O.):=\mathrm{C}^*\left(\{a(\varphi)\mid
                 \varphi\in\ot q.(\al O.) \}\right)
               =\mathrm{CAR}(\ot q.(\al O.),\Gamma\hrist\ot q.(\al O.))
               =\mathrm{CAR}(\overline{\ot q.(\al O.)},
                \Gamma\hrist\overline{\ot q.(\al O.)})
                \subset\mathrm{CAR}(\ot h.,\Gamma)\,.
\]
For the canonical basis projection $P$ given in the context of free nets
(see e.g.~\cite[p.~1157]{Lledo01}) it is also immediate to check that
for double cones
\[
 \ot p.\cap\overline{\ot q.(\al O.)}=
 \ot p.\cap\overline{\ot q.(\al O.)}^\perp=\{0\}\;,\quad\mathrm{where}\;
 \ot p.=P\ot h.\,.
\]
This means that $\ot p.$ and $\overline{\ot q.(\al O.)}$ are in generic 
position and we can apply the results and formulas of the previous and
present sections to the corresponding localized von Neumann algebras
\[
\al M.(\al O.):=\{\pi (a(\varphi))\mid \varphi\in\overline{\ot q.(\al O.)}\}''
                \,. 
\]
In particular from Proposition~\ref{5.3} and
Corollary~\ref{CorDJ} we have that the modular operator $\Delta$ 
and the modular conjugation $J$ are already characterized by their
action on the one-particle Hilbert space. Finally, 
Theorem~\ref{beta*beta} and Remark~\ref{rel1} imply:

\begin{teo}\label{O}
Let $\al O.\subset\R^4$ be a double cone in Minkowski space.
Denote by $Q_\al O.$ the orthoprojection onto $\overline{\ot q.(\al O.)}$
and by $P$ the canonical basis projection given in the context of fermionic
free nets. Then the following formulas hold for the modular operator and
modular conjugation on the one-particle Hilbert space $\ot p.$:
\begin{eqnarray*}
  \mathrm{gra}\,\Delta_\ot p. 
        &=&  \Big\{(PQ_\al O.(p),PQ_\al O.^\perp (p)) \mid p\in\ot p.\Big\}\,.\\
  J(Pq) &=& \Delta_\ot p.^{\frac12}(P\Gamma q)\;,
            \quad q\in\overline{\ot q.(\al O.)}\,.
\end{eqnarray*}
\end{teo}

%%%%%%%%%%%%%%%%%%%%%%%%%%%%%%%%%%%%%%%%%%%%%%%%%%%%%%%%%%%%%%%%%%%%%%%%%%%%%%
%%%%%%%%%%%%%%%%%%%%%%%%%%%%%%%%%%%%%%%%%%%%%%%%%%%%%%%%%%%%%%%%%%%%%%%%%%%%%%
\section{Relation to the real subspace approach}\label{rel2}

In the present section we will make explicit the relation of the
real subspace approach in \cite{Foit83,Wassermann98} to our 
consistent use of complex subspaces in the self-dual approach.

The projection $\al P.$ on the complex Hilbert space $H$ 
with conjugation $\gamma$ in \cite{Wassermann98} corresponds in 
the self-dual approach (where
$\ot h.:=H\oplus H$ and $\Gamma:=\left(\kern-2mm\begin{array}{cc}
                   0 \kern-2mm & \gamma \\[-2mm]
             \gamma  \kern-2mm &  0
       \end{array}\kern-2mm\right)$)
to a diagonal basis projection $\ot P.:=\left(\kern-2mm\begin{array}{cc}
               \al P. \kern-2mm & 0 \\
               0 \kern-2mm & \gamma \al P.^\perp \gamma
       \end{array}\kern-2mm\right)$. Nondiagonal basis projections
are not considered in \cite{Wassermann98}. Further, in this paper
the author extends by second quantization certain mappings on the 
one-particle Hilbert space. He has then to verify that these second 
quantized operators are the modular objects by checking the KMS condition.
(In contrast to that we construct the modular objects 
on the whole antisymmetric Fock space and show that they restrict to the 
$n$-particle space.)

The following aspect is that in \cite{Foit83} (and \cite{Wassermann98})
real-linear closed manifolds $M$ (resp.~$K$) of the one-particle
(complex) space $H$ ($\ot p.$ in our paper) are used, whereas 
here the subalgebras of the `big' fermion algebra
CAR$(\ot h.,\Gamma)$ are characterized by $\Gamma$-invariant complex 
subspaces of the reference space $\ot h.$. The relation between the
two approaches is given by the following observations:
first, the counterpart of the real $M$ in our approach is given by
$P(\mathrm{Re}(\ot q.))$, where $q\in \mathrm{Re}(\ot q.)$
if $\Gamma q=q$. Note that $P(\mathrm{Re}(\ot q.))$ is a real-linear 
submanifold of $\ot p.$ and in general it is {\em not} closed
(see the foregoing considerations). Now we still need to check that
$P(\mathrm{Re}(\ot q.^\perp))$ corresponds in the real subspace approach
of \cite{Foit83} to $iM'$. (Recall that in \cite{Foit83} one defines
$M'$ as the symplectic complement, i.e.~$M':=\{x\in H\mid
\mathrm{Im}\langle x,m\rangle=0\,,\;m\in M\}\supset M^\perp$).
The next result shows that indeed $P(\mathrm{Re}(\ot q.^\perp))$ and
$iM'$ generate the same von Neumann algebra.
\begin{lem}
Put $M:=P(\mathrm{Re}(\ot q.))$. Then 
$P(\mathrm{Re}(\ot q.^\perp))$ is dense in $(iM')$.
\end{lem}
\begin{beweis}
We show first that $P(\mathrm{Re}(\ot q.^\perp))\subseteq (iM')$. 
For any $(q^\perp+\Gamma q^\perp)\in\mathrm{Re}(\ot q.^\perp)$, $q^\perp
\in\ot q.^\perp$, and since 
\[
iM':=\{p\in\ot p.\mid\langle q+\Gamma q,p\rangle+
      \langle p,q+\Gamma q\rangle=0\,,\;q\in\ot q. \}
\]
the inclusion follows from
\begin{eqnarray*}
  \langle q+\Gamma q,P(q^\perp+\Gamma q^\perp)\rangle 
  + \langle P(q^\perp+\Gamma q^\perp) ,q+\Gamma q\rangle
   &=& \phantom{+}\left(\langle q,P q^\perp\rangle 
             + \langle P\Gamma q^\perp,\Gamma q\rangle \right)\\
   & &+\left(\langle \Gamma q,P q^\perp\rangle
             + \langle P\Gamma q^\perp, q\rangle \right)\\
   & &+\left(\langle q,P\Gamma q^\perp\rangle
             + \langle P q^\perp,\Gamma q\rangle \right)\\
   & &+\left(\langle \Gamma q,P \Gamma q^\perp\rangle
             + \langle P\Gamma q^\perp, q\rangle \right)\\
   &=& 0\,,
\end{eqnarray*}
where for the last equation we have used $\Gamma P+P\Gamma=\Gamma$.
Finally, to show the density statement consider $p_0\in iM'$, i.e.
\begin{equation}\label{*}
  \langle q+\Gamma q,p_0\rangle+
      \langle p_0,q+\Gamma q\rangle=0\,,\;q\in\ot q.\,,
\end{equation}
such that $p_0\perp P(q^\perp+\Gamma q^\perp)$ for all 
$q^\perp\in\ot q.^\perp$. Thus 
$\langle p_0,P(q^\perp+\Gamma q^\perp)\rangle=0$, $q^\perp\in\ot q.^\perp$, 
and replacing $q^\perp$ by $iq^\perp$ we also obtain
$\langle p_0,P(q^\perp-\Gamma q^\perp)\rangle=0$, $q^\perp\in\ot q.^\perp$.
Hence $\langle p_0,Pq^\perp\rangle=0$, $q^\perp\in\ot q.^\perp$, and
$p_0\in\ot q.\cap\ot p.$. But according to Eq.~(\ref{*}) we must also
have $\langle p_0+\Gamma p_0,p_0\rangle+
\langle p_0,p_0+\Gamma p_0 \rangle=0$, which implies $p_0=0$.
\end{beweis}

Finally, the conditions $M\cap iM=\{0\}$ and
$M+iM$ dense in $H$ in \cite{Foit83} are equivalent to our conditions 
$\ot q.\cap\ot p.=\{0\}=\ot q.^\perp\cap\ot p.$ and the projections
$P_1$,$P_2$,$P_3$ in \cite[Proposition~1.5]{Foit83}
correspond in the self-dual approach to
the orthoprojections onto $\ot p.\cap\ot q.^\perp$,$\ot p.\cap\ot q.$,
$\ot p.\ominus (\ot p.\cap\ot q.^\perp\oplus\ot p.\cap\ot q.)$,
respectively.

%%%%%%%%%%%%%%%%%%%%%%%%%%%%%%%%%%%%%%%%%%%%%%%%%%%%%%%%%%%%%%%%%%%%%%%%%%%%%%
%%%%%%%%%%%%%%%%%%%%%%%%%%%%%%%%%%%%%%%%%%%%%%%%%%%%%%%%%%%%%%%%%%%%%%%%%%%%%%
\section{Twisted duality. The general case}\label{AbsDuaII}

We are now in a position to give the proof of twisted duality in the most
general situation. Let $(\ot h.,\Gamma)$ be a Hilbert space with
anti-unitary involution $\Gamma$, $P$ {\em any} basis projection
and $\ot q.$ {\em any} closed $\Gamma$-invariant subspace in $\ot h.$,
to which we associate the orthoprojection $Q$. We adapt the arguments
in \cite[p.~735]{Foit83} to the self-dual approach.

Recall the Halmos decompostion 
$\ot h.=\ot h._0\oplus\ot h._1$ given in Eq.~(\ref{Halmos}), where
$\ot h._0=(\ot p.\cap\ot q.)\oplus(\ot p.\cap\ot q.^\perp)\oplus
(\ot p.^\perp\cap\ot q.)\oplus(\ot p.^\perp\cap\ot q.^\perp)$. 
Since $\Gamma(\ot p.\cap\ot q.^\perp)=(\ot p.^\perp\cap\ot q.^\perp)$
and $\Gamma(\ot p.\cap\ot q.)=(\ot p.^\perp\cap\ot q.)$ it is also 
natural to consider the previous decomposition as
\begin{equation}\label{(1)}
  \ot h.=\ot h._{01}\oplus\ot h._{02}\oplus\ot h._1\;,
\end{equation}
where
$\ot h._{01}:=(\ot p.\cap\ot q.^\perp)\oplus\Gamma(\ot p.\cap\ot q.^\perp)$
and $\ot h._{02}:=(\ot p.\cap\ot q.)\oplus\Gamma(\ot p.\cap\ot q.)$.
In particular we have
\begin{eqnarray}
  \label{(2)}
  Q\ot h._{01}=\{0\} &,& Q\ot h._{02}=\ot h._{02} \\  \label{(3)}
  Q^\perp\ot h._{01}=\ot h._{01} &,& Q^\perp \ot h._{02}=\{0\}
\end{eqnarray}
as well as
\[
 \ot q.=\{0\}  \oplus\ot h._{02}\oplus Q\ot h._1
    \quad\mathrm{and}\quad
 \ot q.^\perp=\ot h._{01}  \oplus\{0\}\oplus Q^\perp\ot h._1\,.
\]

\begin{teo} (Twisted Duality)
Let $(\ot h.,\Gamma)$, $P$ and $\ot q.$ be given as in the beginning of
this section. Then
\[
 \al M.(\ot q.)'= \widetilde{Z}\,\al M.(\ot q.^\perp)\, \widetilde{Z}^*\,.
\]
\end{teo}
\begin{beweis}
From Proposition~\ref{TwisCau} it is enough to show the inclusion
\begin{equation}\label{GDInc}
 \al M.(\ot q.)'\subseteq
     \widetilde{Z}\,\al M.(\ot q.^\perp)\, \widetilde{Z}^*\,.
\end{equation}

We can apply now the formulas (\ref{DefTensor}) {\em and} (\ref{DefTensor2})
in Proposition~\ref{Tensor} to 
the 3 space decomposition in Eq.~(\ref{(1)}). Indeed, adapting in the obvious
way the notation from Proposition~\ref{Tensor} we get that
\begin{eqnarray*}
 \lefteqn{\pi\Big(a(f_{01}\oplus f_{02}\oplus f_1)\Big)}\\
   &=& \pi_{01}\Big(a(f_{01})\Big)\otimes Z_{02}\otimes Z_1 +
    \EINS\otimes\pi_{02}\Big(a(f_{02})\Big)\otimes\EINS
    +\EINS\otimes Z_{02}\otimes\pi_1\Big(a(f_1)\Big)\,,
\end{eqnarray*}
$f_1\in\ot h._1$,$f_{0k}\in\ot h._{0k}$, $k=1,2$,
specifies a representation of CAR$(\ot h.,\Gamma)$ on the 
corresponding tensor product
of antisymmetric Fock spaces $\ot F.=\ot F._{01}\otimes\ot F._{02}
\otimes\ot F._1$. Now using (\ref{(2)}) we obtain
\[
 \al M.(\ot q.)=
   \C\EINS\otimes\al L.(\ot F._{02})\otimes\al M.(Q\ot h._1)
\]
hence
\[
 \al M.(\ot q.)'=
   \al L.(\ot F._{01})\otimes\C\EINS\otimes\al M.(Q\ot h._1)'\,.
\]
From (\ref{(3)}) we also obtain
\begin{eqnarray*}
 \al M.(\ot q.^\perp)&=&
   \al L.(\ot F._{01})\otimes\{Z_{02}\}''\otimes\al M.(Q^\perp\ot h._1)\\
   &=&\{ L_{01}\otimes Z_{02}\otimes\pi_1(a(f_1))
        \mid L_{01}\in\al L.(\ot F._{01})\;,\,f_1\in Q^\perp\ot h._1\}''\,.
\end{eqnarray*}
Using now the result stated in Remark~\ref{EOGen} we have for the 
twisted von Neumann algebra
\begin{eqnarray*}
 \widetilde{Z}\al M.(\ot q.^\perp)\widetilde{Z}^*
   &=& \{iZ\pi(a(q^\perp))\mid q^\perp\in\ot q.^\perp\}'' \\
   &=&\{ (Z_{01}\otimes Z_{02}\otimes Z_1)\cdot
        (L_{01}\otimes Z_{02}\otimes\pi_1(a(f_1)))
        \mid L_{01}\in\al L.(\ot F._{01})\;,\,f_1\in Q^\perp\ot h._1\}''\,,
\end{eqnarray*}
which immediately implies (\ref{GDInc}), since we have already proved
twisted duality in the generic position case (cf.~Theorem~\ref{TD0}).
\end{beweis}

%%%%%%%%%%%%%%%%%%%%%%%%%%%%%%%%%%%%%%%%%%%%%%%%%%%%%%%%%%%%%%%%%%%%%%%%%%%%%%
%%%%%%%%%%%%%%%%%%%%%%%%%%%%%%%%%%%%%%%%%%%%%%%%%%%%%%%%%%%%%%%%%%%%%%%%%%%%%%
\section*{Appendix}
We will give in this appendix the proof of Proposition~\ref{Formel}.
Recall the notation and results of Section~\ref{CAR}.\\[3mm]
{\sffamily\bfseries Proposition}
{\em For $f_1,\ldots,f_n\in \ot h.$ the equation
\[
 \Big(a(f_n)\cdot\ldots\cdot a(f_1)\Big)\,\Omega = 
  \sum\limits_{\mbox{\tiny $\begin{array}{c}\pi\in\ot S._{n,\,p}
               \\[1mm] 0\leq 2p\leq n\end{array}$}}
  \!\!\!\!({\rm sgn}\,\pi)\;
  \prod\limits_{l=1}^p \;\langle Pf_{\alpha_l}\,,\,P\Gamma f_{\beta_l}\rangle
  \, P\Gamma f_{j_1} \land \ldots \land P\Gamma f_{j_k} 
\]
holds, where the indices $\alpha_l, \beta_l,j_1,\ldots,j_k$ are given
in the definition of $\ot S._{n,\,p}$ and where for $n=2p$ in the preceding
sum one replaces the wedge product by the vacuum $\Omega$.}\\[2mm]
\begin{beweis}
The proof is done by induction on the number of generators of the 
CAR-Algebra. For $n=1$ the above formula is immediately verified
using the definition of creation and annihilation operators.
Assume that it holds for $n$ generators 
and we prove that it is also true for $n+1$ generators. Take
$f_{n+1},f_n,\ldots,f_1\in\ot h.$ and from the preceding 
assumption as well as the results stated in Section~\ref{CAR} we have
\begin{eqnarray}
\lefteqn{\!\!\!\!\!\!\!\!\!\!\!\!\!\!\!\!\!\!\!\!\!
     a(f_{n+1})\,\Big(a(f_n)\cdot\ldots\cdot a(f_1)\,\Omega\Big)}
     \nonumber\\[2mm]
           &=& \Big(c(P\Gamma f_{n+1})^*+c(Pf_{n+1})\Big)\; 
     \Big((a(f_n)\cdot\ldots\cdot a(f_1)\,\Omega\Big) \nonumber \\[2mm]
\label{T1} &=& \sum\limits_{\mbox{\tiny $\begin{array}{c}\pi\in\ot S._{n,\,p}
               \\[1mm] 0\leq 2p\leq n\end{array}$}}
  \!\!\!\!({\rm sgn}\,\pi)\;
  \prod\limits_{l=1}^p \;\langle Pf_{\alpha_l}\,,\,P\Gamma f_{\beta_l}\rangle
  \,P\Gamma f_{n+1}\land P\Gamma f_{j_1}\land\ldots\land P\Gamma f_{j_k}\\ 
 & & +\sum\limits_{\mbox{\tiny $\begin{array}{c}\pi\in\ot S._{n,\,p}
               \\[1mm] 0\leq 2p\leq n\end{array}$}}
    \sum_{r=1}^k \;\;({\rm sgn}\,\pi)\, (-1)^{r-1}\;
  \langle Pf_{n+1}\,,\,P\Gamma f_{j_r}\rangle\cdot \nonumber\\
 & & \;\;\;\;\;\cdot\prod\limits_{l=1}^p 
  \;\langle Pf_{\alpha_l}\,,\,P\Gamma f_{\beta_l}\rangle
  \,P\Gamma f_{j_1}\land\ldots \land\widehat{P\Gamma f_{j_r}}
  \land\ldots\land P\Gamma f_{j_k}\,. \nonumber
\end{eqnarray}
We will determine how many terms with particle number $k'$ appear in the
preceding sum. For this let $p'\in\N$ be such that $2p'+k'=n+1$. Now
the first term in the above formula contributes by means of expressions
where $k=k'-1$ (hence $p=p'$) and there are 
\[
 \left(\kern-1.5mm\begin{array}{c}
 n \\ n-2p
 \end{array}\kern-1.5mm\right)
 \frac{(2p)!}{p!\,2^p}
=\left(\kern-1.5mm\begin{array}{c}
 n \\ n-2p'
 \end{array}\kern-1.5mm\right)
 \frac{(2p')!}{p'!\,2^{p'}}
\]
such summands. Further, the second term contributes by means of
expressions where $k=k'+1$ (hence $p=p'-1$) and there are now
\[
 k\,\left(\kern-1.5mm\begin{array}{c}
     n \\ n-2p
    \end{array}\kern-1.5mm\right)
    \frac{(2p)!}{p!\,2^p}
=   \left(\kern-1.5mm\begin{array}{c}
     n \\ n-2p'+1
    \end{array}\kern-1.5mm\right)
    \frac{(2p')!}{p'!\,2^{p'}}
\]
such summands. Altogether we obtain
\[
 \left(\kern-1.5mm\begin{array}{c}
 n \\ n-2p'
 \end{array}\kern-1.5mm\right)
 \frac{(2p')!}{p'!\,2^{p'}}
+ \left(\kern-1.5mm\begin{array}{c}
     n \\ n-2p'+1
    \end{array}\kern-1.5mm\right)
    \frac{(2p')!}{p'!\,2^{p'}}
= \left(\kern-1.5mm\begin{array}{c}
     n+1 \\ n+1-2p'
    \end{array}\kern-1.5mm\right)
    \frac{(2p')!}{p'!\,2^{p'}}
\]
terms with particle number $k'$ and this coincides with the number of
elements in $\ot S._{n+1,\,p'}$. To conclude the proof we still need to show
that each term in the sum carries the correct sign. For the summands
in (\ref{T1}) this follows from
\begin{eqnarray*}
\lefteqn{\!\!\!\!\!\!\!\!\!\!\!\!\!\!\!\!\!\!\!\!\!\!\!\!\!\!\!\!\!\!\!
       {\rm sgn}\mbox{\scriptsize
       $\left(\kern-1.5mm {\begin{array}{cccccccc}
       n & n-1     &\cdots & n-2p+2  & n-2p+1 & k   & \cdots & 1 \\
\alpha_1 &\beta_1 &\cdots &\alpha_p &\beta_p & j_1 & \ldots & j_k   
          \end{array}}\kern-1.5mm\right)$}}   \\[2mm] 
 &=& {\rm sgn}\mbox{\scriptsize
       $\left(\kern-1.5mm {\begin{array}{cccc}
       n+1 & n         &\cdots &  1 \\
       n+1 & \alpha_1  &\cdots &  j_k   
          \end{array}}\kern-1.5mm\right)$} \\[2mm]
 &=& {\rm sgn}\mbox{\scriptsize
     $\left( \kern-1.5mm{\begin{array}{ccccccccc}
     n+1 & n      &\cdots & n-2p+3  & n-2p+2 & k+1 & k   & \cdots & 1 \\
\alpha_1 &\beta_1 &\cdots &\alpha_p &\beta_p & n+1 & j_1 & \ldots & j_k   
          \end{array}}\kern-1.5mm\right)\;$}.
\end{eqnarray*}
For the remaining terms we have 
\begin{eqnarray*}
\lefteqn{\!\!\!\!\!\!\!\!\!\!\!\!\!\!\!\!\!\!\!\!\!\!\!\!\!\!\!\!\!\!\!\!\!
    (-1)^{r-1}\;{\rm sgn}\mbox{\scriptsize
       $\left(\kern-1.5mm {\begin{array}{cccccccccc}
      n & n-1   &\cdots & k   & \dots & k+2-r  & k+1-r & k-r &\cdots & 1 \\
\alpha_1&\beta_1&\cdots & j_1 & \dots & j_{r-1}& j_r   & j_{r+1}&\ldots & j_k 
          \end{array}}\kern-1.5mm\right)$}}   \\[2mm] 
 &=& (-1)^{r-1}\;{\rm sgn}\mbox{\scriptsize
       $\left(\kern-1.5mm {\begin{array}{cccc}
       n+1 & n         &\cdots &  1 \\
       n+1 & \alpha_1  &\cdots &  j_k   
          \end{array}}\kern-1.5mm\right)$} \\[2mm]
 &=& {\rm sgn}\mbox{\scriptsize
     $\left( \kern-1.5mm{\begin{array}{ccccccccccc}
     n+1& n & n-1    & n-2   &\cdots& k &\cdots&k+1-r  &k-r &\cdots & 1 \\
     n+1&j_r&\alpha_1&\beta_1&\cdots&j_1&\cdots&j_{r-1}&j_{r+1}&\ldots&j_k   
          \end{array}}\kern-1.5mm\right)\;$}.
\end{eqnarray*}
Therefore we have shown that
\[
 \Big(a(f_{n+1})\cdot\ldots\cdot a(f_1)\Big)\,\Omega = 
  \sum\limits_{\mbox{\tiny $\begin{array}{c}\pi\in\ot S._{n+1,\,p'}
               \\[1mm] 0\leq 2p'\leq n+1\end{array}$}}
  \!\!\!\!({\rm sgn}\,\pi)\;
  \prod\limits_{l=1}^{p'} 
  \;\langle Pf_{\alpha_l}\,,\,P\Gamma f_{\beta_l}\rangle
  \, P\Gamma f_{j_1} \land \ldots \land P\Gamma f_{j_k}
\]
which concludes the proof.
\end{beweis}

\paragraph{Acknowledgements}
It is a pleasure to thank Dr.~H.~Neidhardt for discussions 
on the subject, in particular on Proposition~\ref{Fall=1}
and Theorem~\ref{Equivalence}. We would also like to 
acknowledge the remark of a referee (concerning
Section~\ref{AbsDuaII}) that pointed at the arguments in
\cite[p.735]{Foit83}. 
Finally, one of us (F.Ll.) wants to thank Dr.~C.~Binnenhei 
for a useful conversation in G\"ottingen.

%\bibliographystyle{amsplai1}
%\bibliography{qft}

\providecommand{\bysame}{\leavevmode\hbox to3em{\hrulefill}\thinspace}

\end{document}